\documentclass[conference]{IEEEtran}
\IEEEoverridecommandlockouts
\usepackage{cite}
\usepackage{amsmath}
\usepackage{algorithmic}
\usepackage{algorithm}  
\usepackage{graphicx}
\usepackage{textcomp}
\usepackage{xcolor}
\usepackage{float} 
\usepackage{bm}
\usepackage{multirow}
\usepackage{dblfloatfix} 
\usepackage{array}
\usepackage{booktabs}
\usepackage{fancyhdr} 
\usepackage{color}
\usepackage{tabu}
\usepackage{nomencl}
\makenomenclature
\usepackage{etoolbox}
\usepackage{graphicx}
\usepackage{subfigure}
\usepackage{amsfonts,amssymb}
\usepackage{mathrsfs}
\usepackage[colorlinks, citecolor=black]{hyperref}
\hypersetup{ colorlinks=true, linkcolor=black, citecolor=black, urlcolor = black }

\usepackage{nomencl}
\makenomenclature
\newcommand{\nomenclaturewww}[2]{\nomenclature{\textcolor{black}{#1}}{\textcolor{black}{#2}}}

\newcolumntype{I}{!{\vrule width 1.3pt}}
\newlength\savedwidth

\newlength\savewidth

\def\BibTeX{{\rm B\kern-.05em{\sc i\kern-.025em b}\kern-.08em
    T\kern-.1667em\lower.7ex\hbox{E}\kern-.125emX}}

\newcommand{\blue}[1]{\textcolor{black}{{#1}}}

\begin{document}

\title{Cooperative Hierarchical Deep Reinforcement Learning based Joint Sleep and Power Control in RIS-aided Energy-Efficient RAN \\
}

\author{\IEEEauthorblockN{Hao Zhou, Medhat Elsayed, Majid Bavand, Raimundas Gaigalas,
Steve Furr, \\Melike Erol-Kantarci, \IEEEmembership{Senior Member, IEEE} \vspace{-20pt} }

\thanks{ This work has been supported by MITACS and Ericsson Canada, and NSERC Collaborative Research and Training Experience Program (CREATE) under Grant 497981. This article was presented in part at the IEEE GLOBECOM 2022 \cite{mel1}.

H. Zhou and M. Erol-Kantarci are with the School of Electrical Engineering and Computer Science, University of Ottawa, Ottawa, ON K1N 6N5, Canada. (email:\{hzhou098, melike.erolkantarci\}@uottawa.ca) 

M. Elsayed and M. Bavand, are with Ericsson Canada, ON K2K 2V6, Ontario, Canada. (email:\{medhat.elsayed,majid.bavand\}@ericsson.com)

S. Furr is with Amazon, USA. The work was done while he was at Ericsson Canada. (email: furrste@amazon.com)

R. Gaigalas is with Ericsson Sweden, 164 80, Stockholm County, Sweden. (email:\{raimundas.gaigalas\}@ericsson.com)
}}

\maketitle

\thispagestyle{fancy}            
\chead{This paper has been accepted by IEEE Transactions on Cognitive Communications and Networking.} 

\renewcommand{\headrulewidth}{0pt}      
\pagestyle{plain}

\begin{abstract}
Energy efficiency (EE) is one of the most important metrics for envisioned 6G networks, and sleep control, as a cost-efficient approach, can significantly lower power consumption by switching off network devices selectively.
Meanwhile, the reconfigurable intelligent surface (RIS) has emerged as a promising technique to enhance the EE of future wireless networks. 
In this work, we jointly consider sleep and transmission power control for RIS-aided energy-efficient networks.
In particular, considering the timescale difference between sleep control and power control, we introduce a cooperative hierarchical deep reinforcement learning (Co-HDRL) algorithm, enabling hierarchical and intelligent decision-making. Specifically, the meta-controller in Co-HDRL uses cross-entropy metrics to evaluate the policy stability of sub-controllers, and sub-controllers apply the correlated equilibrium to select optimal joint actions. Compared with conventional HDRL, Co-HDRL enables more stable high-level policy generations and low-level action selections.   
Then, we introduce a fractional programming method for RIS phase-shift control, maximizing the sum-rate under a given transmission power. In addition, we proposed a low-complexity surrogate optimization method as a baseline for RIS control. 
Finally, simulations show that the RIS-assisted sleep control can achieve more than 16\% lower energy consumption and 30\% higher EE than baseline algorithms.
\end{abstract}



\mbox{}
\nomenclaturewww{$\bm{H}_{b,m}$}{The link between base station $b$ and RIS $m$}
\nomenclaturewww{$g_{b,m}$}{The path loss from $b^{th}$ BS to $m^{th}$ RIS}
\nomenclaturewww{$\mathcal{N}_m$}{The total number of elements of $m^{th}$ RIS}
\nomenclaturewww{$\bm{h}_{b,m,n}$}{The phase difference in $\bm{H}_{b,m}$}
\nomenclaturewww{$d_{b,m,n}$}{The distance from $b^{th}$ BS to $m^{th}$ RIS}
\nomenclaturewww{$\lambda$}{The signal wavelength}
\nomenclaturewww{$\text{diag}$}{The diagonal operator}
\nomenclaturewww{$\omega_{m}$}{The amplitude reflection coefficient of $m^{th}$ RIS}
\nomenclaturewww{$\theta_{m,n}=e^{j\theta'_{m,n}}$}{ The phase shift of $n^{th}$ RIS elements}
\nomenclaturewww{$\bm{\Theta}_{m}$}{The phase shift matrix of RIS $m$}
\nomenclaturewww{$\mu$}{The resolution of the RIS element’s phase shifter}
\nomenclaturewww{$\tau_{m,k}$}{The Rician factor between $m^{th}$ RIS and $k^{th}$ UE}
\nomenclaturewww{$\overline{\bm{G}}_{m,k}$, $\widetilde{\bm{G}}_{m,k}$}{Deterministic line-of-sight and non-line-of-sight components in Rician fading channel}
\nomenclaturewww{$C_{b,k}$}{The achievable rate between BS $b$ and UE $k$}
\nomenclaturewww{$b_{k}$}{The bandwidth allocated to user $k$}
\nomenclaturewww{$N_{0}^2$}{The noise power}
\nomenclaturewww{$\mathcal{K}_{b}$}{The set of users in BS $b$}
\nomenclaturewww{$p_{b,k}$}{The transmission power of $b^{th}$ BS for user $k$}
\nomenclaturewww{$\mathcal{B}$}{The set of all BSs}
\nomenclaturewww{$\mathcal{M}$}{The set of RISs }
\nomenclaturewww{$E_{b}$}{The energy consumption of $b^{th}$ BS}
\nomenclaturewww{$q_b$}{The sleep mode indicator. $q_b=1$ indicates active mode, and $q_b=0$ means sleep mode}
\nomenclaturewww{$P_{active}$}{Constant energy consumption of BSs in active mode}
\nomenclaturewww{$\delta_{BS}$}{Slope of load-dependent energy consumption factor}
\nomenclaturewww{$P_{sleep}$}{The BS energy consumption in sleep mode}
\nomenclaturewww{$P_{b}$}{The total transmission power of $b^{th}$ BS}
\nomenclaturewww{$P_{b,max}$}{The maximum transmission power of BS $b$}
\nomenclaturewww{$S,A,R,T$}{Set of states, actions, reward and transition probability in Markov decision process }
\nomenclaturewww{$\mathcal{G}$}{The set of goals}
\nomenclaturewww{$W_{b,k}$}{The transmission demand of UE $k$ in BS $b$}
\nomenclaturewww{$s_{sub}$, $a_{sub}$, $r_{sub}$}{State, action, and reward of sub-controllers}
\nomenclaturewww{$s_{meta}$, $g_{meta}$, $r_{meta}$}{State, goal, and reward of the meta-controllers}
\nomenclaturewww{$Er$}{ The loss function of neural network training }
\nomenclaturewww{$w$,$w'$}{The weight of the main and target networks}
\nomenclaturewww{$\gamma$}{The discount factor}
\nomenclaturewww{$I(X)$}{The entropy of $x$ in set $X$}
\nomenclaturewww{$D_{KL}(X||Y)$}{Kullback–Leibler divergence to define the relative entropy from one distribution $X$ to another distribution $Y$}
\nomenclaturewww{$N^{X}$}{The total number of possible outcomes of variable $x$ in a set $X$}
\nomenclaturewww{$pr(x_{i})$}{The probability of $x_{i}$ in set $X$}
\nomenclaturewww{$\beta$, $\sigma_{m,n}$}{Auxiliary variables given by Lagrangian dual transform}
\nomenclaturewww{$\psi_{k}$}{The signal-to-interference-plus-noise ratio of user $k$}
\nomenclaturewww{$\bm{\eta}$, $\bm{V}_{b,k}$}{Collections of auxiliary variables in the problem transformation }
\nomenclaturewww{$\text{Re}\{\}$}{The real number part of a complex number}
\nomenclaturewww{$W^{max}_{b}$}{The max transmission demand of BS $b$}
\nomenclaturewww{$z_{m,n}$}{Discrete RIS phase-shift selections.}
\nomenclaturewww{}{}
\nomenclaturewww{}{}
\nomenclaturewww{}{}
\printnomenclature

\section{Introduction}
Energy efficiency (EE) is a critical metric for sustainable 5G and 6G networks. Sleep control has been one of the widely considered approaches to enhance the EE of radio access networks (RANs), which turns network devices such as base stations (BS) to sleep mode selectively \cite{b2}. On the other hand, the reconfigurable intelligent surface (RIS) has emerged as a promising technique for 5G beyond and 6G networks. In particular, a huge number of small and low power consumption elements are integrated into RISs where each can forward the incoming signal. Since no power amplifier is needed, RISs consume much less energy than conventional relay transceivers. Therefore, RISs can become an ideal solution to improve the EE of RAN \cite{b3}.
However, these potential solutions for energy-efficient RAN may greatly increase the network management complexity as multiple network functions should be jointly considered to adapt to network dynamics. Fortunately, machine learning techniques offer promising opportunities for intelligent network management and control, and the advantage of machine learning algorithms has been extensively studied in many works \cite{b34}.  

In this work, we jointly consider sleep control, transmission power control and RIS control to improve EE performance. The BS transmission power control is involved because: i) if one BS enters the sleep mode, the rest active BSs may need to increase the transmission power to serve the user equipment (UE) that is previously associated with the BS that has switched to sleep mode; ii) BS transmit power and RIS phase shifts are usually jointly optimized, which is known as joint active and passive beamforming\cite{b10}; iii) power control can reduce the overall interference level, increase the average Signal-to-interference-plus-noise ratio (SINR), and further improve the EE\cite{b5}. However, compared with transmit power and RIS control, sleep control is a long-term decision that affects the network performance for several consecutive time slots. 
By contrast, power control applies faster decision-making to adapt to the changing traffic demand of UEs. Therefore, such timescale obstacles in decision-making prevent the application of conventional reinforcement learning algorithms \cite{b6}. 

To this end, we deploy a hierarchical architecture for joint sleep and power control, including a meta-controller for long-term sleep control and several sub-controllers for short-term power control. The sleep control is decided every few time slots as a high-level policy for sub-controllers, and sub-controllers can adjust the transmission power in each time slot accordingly.
HRL has been used for joint relay selection and power optimization in \cite{b7} and multichannel sensing in \cite{b7-1}. But in the existing schemes, the meta-controller uses the average reward of sub-controllers as feedback to assess the high-level policy. Nevertheless, the low-level action selection is constantly changing, and the average reward metric may have difficulty representing the sub-controller status. 
Learning multiple levels of policies simultaneously can be difficult due to the inherent instability. Specifically, changes in a policy at one level of the hierarchy may lead to changes in the policies of higher levels, making it difficult to jointly learn multiple levels of policies. For instance, a high average reward may be produced by very unstable sub-controllers that cannot guarantee a satisfying performance. Such unstable low-level policies can mislead the decision-making of higher-level controllers, which has become a critical issue for HRL applications \cite{b8}. 

Therefore, we propose a novel cooperative hierarchical deep reinforcement learning (Co-HDRL) algorithm. For high-level meta-controller, we propose a cross-entropy enabled policy to monitor the stability of sub-controllers, which aims at evaluating the reliability of low-level reward feedback. In particular, cross-entropy can evaluate the stability of low-level action selections under given high-level policies, and then provide more reliable feedback about the sub-controllers' performance. Therefore, the meta-controller can estimate the optimality of its policies more accurately.
For low-level sub-controllers, we introduce a correlated equilibrium-based cooperation strategy to stabilize action selections. With correlated equilibrium, multiple sub-controllers can make the optimal joint actions that will benefit the whole system.  
Therefore, Co-HDRL is expected to produce a stable performance under a hierarchical scheme while allowing a long-term RAN policy to interact with short-term optimizations of network functions.

On the other hand, compared with sleep and power control, RIS phase-shift control has more stringent requirements for fast and real-time optimization \cite{b1}. Each RIS element should respond to the incoming signal immediately, and the channel condition may be randomly changed in the next time slot, which requires an optimization algorithm with rapid responses.
\blue{There have been many advanced algorithms for RIS configuration. For example, low-complexity channel estimation and passive beamforming designs are proposed in \cite{an2021low} for rapidly time-varying channels, and a codebook-based framework is proposed in \cite{an2022codebook} to investigate the trade-offs between communication performance and signalling overhead.} 
\blue{Machine learning-based algorithms also provide promising solutions for RIS control, e.g., deep deterministic policy gradient (DDPG) is applied in \cite{xu2022deep} to handle continuous action spaces for joint beamforming in a location-aware imitation environment.}
\blue{Meanwhile, convex optimization is the most widely used optimization technique, which has been extensively investigated in many studies to solve RIS-related optimization problems\cite{zhou2023survey}, i.e., utilizing fractional programming (FP) techniques to maximize the weighted sum-rate \cite{b11}.} 
In this work, we focus more on hierarchical learning-based decision-making, and it is reasonable to deploy a well-developed algorithm for RIS phase-shift control.
\blue{Compared with machine learning algorithms such as DDPG, convex optimization can provide more stable outputs by achieving a closed-form output. In our proposed hierarchical architecture, a stable low-level RIS controller can also provide more reliable output to the high-level meta-controller, and then we can better focus on the stability investigation of the defined hierarchical architecture.
Therefore, we apply an FP-based algorithm to maximize the sum-rate of all UEs, which has been successfully used in many existing studies and achieves closed-form solutions \cite{b9}, e.g, using FP to solve sum-of logarithmic-ratios problems in \cite{xu2023algorithm}.}

The main contribution of this work is that we propose a novel Co-HDRL algorithm for intelligent hierarchical decision-making in wireless networks to handle problems with different timescales. More specifically, a cross-entropy-enabled policy is proposed for the high-level meta-controller to evaluate the low-level stability, and correlated equilibrium is deployed for sub-controllers to stabilize the joint action selections. 
Moreover, such joint control capabilities for handling timescale differences are critical to many existing platforms such as O-RAN and associated control loops. 
The simulations show that Co-HDRL can achieve 16\% lower power consumption and 35\% higher EE than baseline algorithms. Moreover, Co-HDRL presents a more stable action selection policy and higher rewards than conventional HDRL algorithms.

The rest of this work is organized as follows. Section \ref{s2} presents the related work, and Section \ref{s3} gives the network and system model. Section \ref{s5} introduces the proposed Co-HDRL algorithm, and the RIS phase-shift optimization methods are presented in Section \ref{s4}. Finally, Section \ref{s7} shows the simulation results, and Section \ref{s8} concludes this work.

\section{Related Work} \label{s2}
Machine learning techniques, including Q-learning\cite{b12}, deep Q-learning, transfer learning, and actor-critic learning\cite{b15} have been widely used for wireless network management. For example, Q-learning and deep Q-learning are used in \cite{b14} for energy sharing of renewable energy powered base stations to reduce energy footprint. \cite{b15} combines actor-critic reinforcement learning with spatial-temporal networks for traffic-aware BS sleep control. \cite{b33} defined a deep Q-learning-based scheme with action-wise experience replay and adaptive reward scaling for sleep control. However, it is worth noting that most existing works require actions with the same timescales, which means all the actions have to be decided simultaneously. But in practice, the actions may belong to different agents that apply various decision-making timescales. 

Consequently, HRL is proposed to overcome this issue \cite{mel2}. \cite{b7} proposed an HRL-based method for joint relay selection and power optimization in two-hop cooperative relay networks to reduce outage probability. In \cite{b7-1}, HRL is used for dynamic spectrum access control, which aims to reduce the complexity of the whole optimization process.
In these works, the meta-controller uses the average reward of sub-controllers to evaluate the high-level policy. But the low-level stability issue is not considered since the low-level action selection is constantly changing\cite{b8,b16}. Then the unreliable reward feedback from sub-controllers may mislead the policy selection of meta-controllers and degrade the overall performance. In our former work \cite{mel1}, we provide a hierarchical Q-learning-based method for joint sleep and power control, but RIS control is not included in our prior work. In addition, the sub-controller stability is not considered in \cite{mel1} by using conventional $\epsilon$-greedy policy for the action and goal selection for the sub-controller and meta-controller, which may lead to unstable policy selection.
To this end, this work proposes a Co-HDRL algorithm to overcome this issue, including a cross-entropy-enabled meta-controller policy to monitor the low-level stability, and a correlated equilibrium to stabilize the action selection of sub-controllers. 

Compared with existing sleep control studies, another difference is that we investigate how the RIS technique can contribute to sleep control. RISs have become a promising technique to improve the EE of 5G beyond and 6G networks. For instance, \cite{b17} proposed two algorithms for the transmit power allocation and the phase shifts of RIS elements to maximize the EE metric, and \cite{b18} studied the tradeoff between energy efficiency and spectral efficiency in RIS-aided networks. 
RIS phase-shift optimization has been extensively studied, and therefore this work applies a well-known FP-based method for RIS phase-shift optimization. Note that many other algorithms can be used to optimize RIS phase shifts, and here we select the FP-based method since it has been widely used in the literature. 

\section{Network and System Model}\label{s3}

\begin{figure}[!t]
\centering
\includegraphics[width=0.95\linewidth]{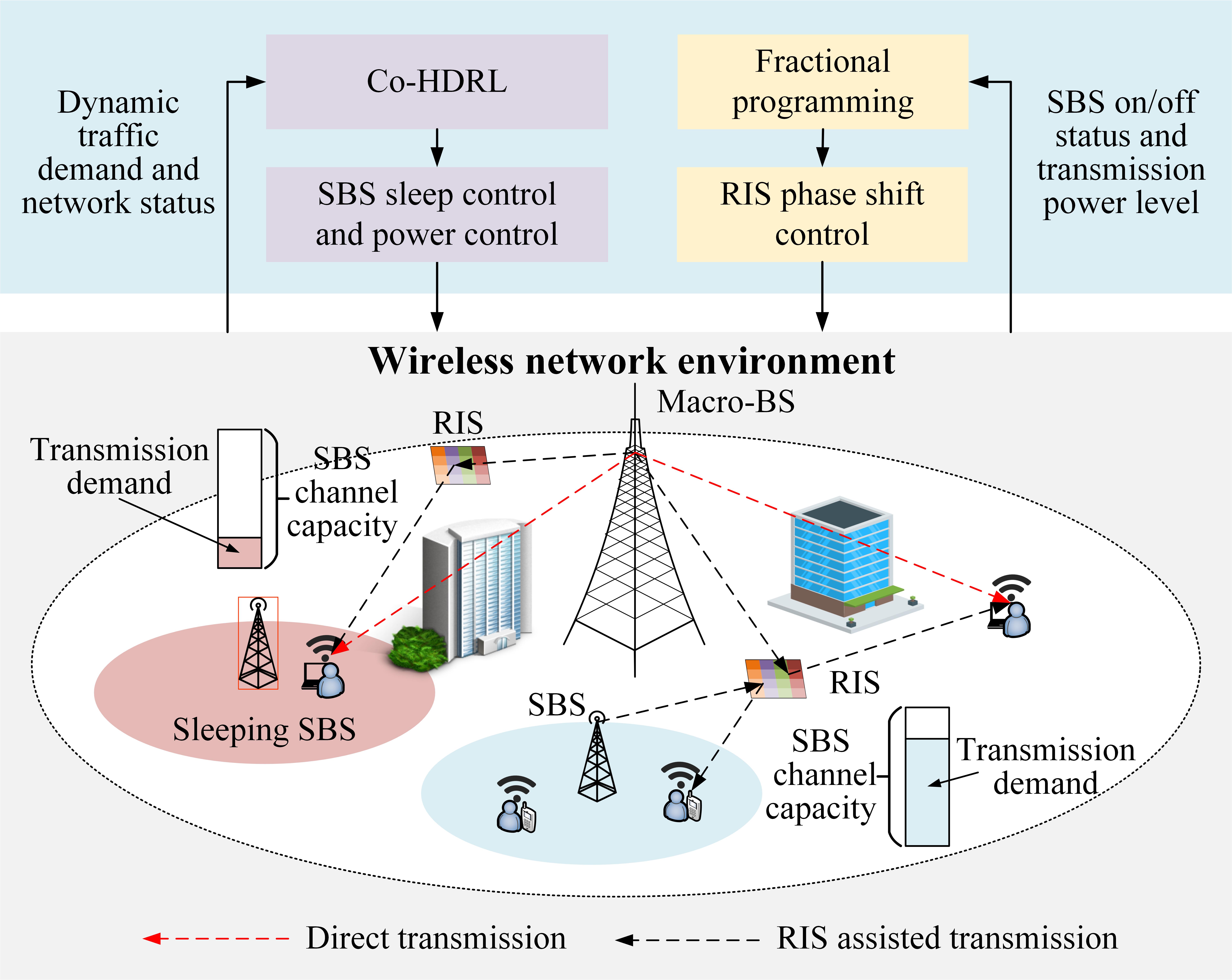}
\caption{RIS-aided heterogeneous network.}
\label{fig1}
\setlength{\abovecaptionskip}{-2pt} 
\vspace{-10pt}
\end{figure}

As shown in Fig. \ref{fig1}, we consider a multi-BS environment. Firstly, the small base stations (SBSs) can enter sleep mode when the traffic demand drops, which will reduce energy consumption and increase overall EE. Then the \blue{macro base station (MBS)} will take over the active UEs that are previously covered by sleeping SBSs\footnote{\blue{ We use "BS" to represent all BSs, including both SBSs and MBS. MBS and SBS are illustrated in Fig. \ref{fig1}.}}. 
Meanwhile, SBSs can adjust their transmission power dynamically to achieve the desired data rate for attached UEs and reduce interference on other SBSs. 
In this work, we propose a Co-HDRL method for joint sleep and power control of SBSs.
On the other hand, the direct transmission between BSs and UEs may suffer high penetration loss due to dense buildings. Then, RISs are deployed to reshape the signal transmission path and increase the SINR, and we apply an FP-based algorithm for the RIS phase-shift control.
\blue{Finally, for the BS-RIS-UE associations, we assume each SBS has a fixed coverage area, and UEs that are not covered by any SBSs will be associated with the MBS. In addition, one RIS can be simultaneously used by multiple BSs to serve different users, which is a common setting in many existing association-related studies\cite{liu2022joint}.}

\subsection{RIS-Assisted Channel Model}
The direct transmissions between BS and UEs are usually considered as non-line-of-sight (NLOS) with a high penetration loss caused by dense buildings and complicated propagation environments. Then, we assume UEs mainly receive signals by the indirect RIS-assisted transmission that consists of BS-RIS and RIS-UE links\cite{b17}.

RIS elements are usually deployed on the surface of high buildings, and then the BS-RIS link is considered as line-of-sight (LOS) link 
$ \bm{H}_{b,m}=g_{b,m}[\bm{h}_{b,m,1},...,$ $\bm{h}_{b,m,n},...,\bm{h}_{b,m,|\mathcal{N}_m|}]$,
where $g_{b,m}$ is the path loss from $b^{th}$ BS to $m^{th}$ RIS, $\mathcal{N}_m$ is the total number of elements of $m^{th}$ RIS. The phase difference $\bm{h}_{b,m,n}$ is given by $\bm{h}_{b,m,n} = \exp\left(\frac{{-2j\pi d_{b,m,n}}}{\lambda}\right)$,
where $j$ is the imaginary unit, $d_{b,m,n}$ is the distance from $b^{th}$ BS to $m^{th}$ RIS, $\lambda$ is the signal wavelength.

Then, the signal will be reflected by RISs to UEs by a phase shift matrix:
$ \bm{\Theta}_{m}=\omega_{m}\text{diag}(\theta_{m,1},...,\theta_{m,n},...,\theta_{m,\mathcal{N}_m})$,
where $\text{diag}$ indicates the diagonal operator, $\omega_{m}$ is the amplitude reflection coefficient of $m^{th}$ RIS, and $\omega_{m}\in [0,1]$. $\theta_{m,n}=e^{j\theta'_{m,n}}$ is the phase shift of $n^{th}$ RIS elements with $\theta'_{m,n} \in \left\{0,\frac{2\pi}{2^\mu},\cdots,\frac{2\pi (2^\mu -1)}{2^\mu} \right\}$, and $\mu$ is the resolution of the RIS element’s phase shifter.
Considering the complex environment on the UE side, the RIS-UE link follows the Rician fading, which is a combination of LOS and NLOS transmissions\cite{b19}: 
\begin{equation}\label{eq5}
\bm{G}_{m,k}=\sqrt{\frac{\tau_{m,k}}{\tau_{m,k}+1}}\overline{\bm{G}}_{m,k}+\sqrt{\frac{1}{\tau_{m,k}+1}}\widetilde{\bm{G}}_{m,k},   
\end{equation}
where $\tau_{m,k}$ is the Rician factor between $m^{th}$ RIS and $k^{th}$ UE. $\overline{\bm{G}}_{m,k}$ is the deterministic LOS component. $\widetilde{\bm{G}}_{m,k}$ is the NLOS component, which is given by independent and identically complex Gaussian distribution.
%
%
For a downlink transmission between BS $b$ and UE $k$, the achievable rate is

\begin{equation}\label{eq7}
\blue{
\begin{aligned}
C_{b,k}= b_{k}\log \bigg ( 1+ \qquad \qquad \qquad \qquad \qquad \qquad \qquad \\ 
\frac{|\sum\limits_{m\in \mathcal{M}}\bm{H}_{b,m}\bm{\Theta}_{m}\bm{G}_{m,k}^{\dag}p_{b,k}|^2}{\sum\limits_{b'\in \mathcal{B}} \sum\limits_{k'\in \mathcal{K}_{b'},k'\neq k}|\sum\limits_{m'\in \mathcal{M}}\bm{H}_{b',m'}\bm{\Theta}_{m'}\bm{G}_{m',k}^{\dag}p_{b',k'}|^2+N_{0}^2} \bigg),
\end{aligned}}
\end{equation}
where \blue{$\mathcal{B}$ is the set of BSs\cite{han2021joint},} $b_{k}$ is the bandwidth allocated to user $k$, $\bm{G}_{m,k}^{\dag}$ represents the conjugate transpose of vector $\bm{G}_{m,k}$, $N_{0}^2$ is the noise power, $\mathcal{K}_{b}$ is the set of users in BS $b$, \blue{$p_{b,k}$ is the transmission power} of $b^{th}$ BS for user $k$, and $\mathcal{M}$ is the set of RISs in the environment. \blue{This work considers BS sleep control, and therefore $p_{b,k}$ in equation (\ref{eq7}) will become 0 if the BS $b$ enters sleep mode, which will be introduced in the next section.}

\subsection{BS Energy Consumption Model}
\blue{The energy consumption of the $b^{th}$ \blue{BS} is\cite{b20}}:
\begin{equation} \label{eq8}
 E_{b}=\left\{
\begin{array}{lcl} P_{active}+ \blue{\delta_{BS} P_{b}},     &   \text{if\,} q_b=1, \,\, \text{\blue{for all BSs}};  \\
P_{sleep} ,    &   \text{if\,} q_b=0, \,\, \text{\blue{for SBS only}.}
\end{array} \right.
\end{equation}
where $q_b=1$ indicates active mode, and $q_b=0$ means sleep mode. 
\blue{We assume only SBSs can enter the sleep mode, while the MBS is always in active mode $q_b=1$, providing connections to UEs that are not covered by any SBSs.}
$P_{active}$ is the constant energy consumption of BSs in active mode, $\delta_{BS}$ is the slope of load-dependent energy consumption factor, and $P_{sleep}$ is the BS energy consumption in sleep mode.
\blue{For ease of notation, we define $P_{b}$ to represent total transmission power of $b^{th}$ BS with $P_{b}=\sum_{k \in \mathcal{K}_{b}}||p_{b,k}||$ in equation (\ref{eq8}), where $p_{b,k}$ has been defined in equation (\ref{eq7}).}

\subsection{Problem Formulation}
\blue{The overall objective is to maximize the total EE of MBS and SBSs, and the problem formulation is given by:}
\begin{subequations}\label{e2:main}
\begin{align}
\max\limits_{q_b,p_{b,k},\bm{\Theta}_{m}} & \frac{\sum_{b\in \mathcal{B} } \sum_{k \in \mathcal{K}_{b}} C_{b,k}}{\sum_{b\in \mathcal{B} } E_{b}}  & \tag{\ref{e2:main}} \\
 \text{s.t.}  \quad & \blue{\text{Equation\,} (\ref{eq7}), (\ref{eq8})} & \label{e2:c}  \\ 
& |\theta_{m,n}|^2=1, m\in\mathcal{M}, n\in\mathcal{N}_m, & \label{e2:d} \\
& \blue{P_{b} \leq P_{b,max}}
\end{align}
\end{subequations}
where \blue{$C_{b,k}$ is the achievable rate of $k^{th}$ UE as defined in previous equation (\ref{eq7}), and $E_{b}$ is the total power consumption of $b^{th}$ BS as defined in equation (\ref{eq8}).} 
\blue{$P_{b}$ has been defined in the above subsection as the total transmission power of $b^{th}$ BS}, and $P_{b,max}$ is the maximum transmission power of BS $b$. 
The power consumption of RIS elements is not included here because it is much lower than BS power in practice\cite{b17}.
The defined problem (\ref{e2:main}) includes three control variables $q_b$, $P_{b}$, and $\bm{\Theta}_{m}$. First, the binary variable $q_b$ in \blue{the equation (\ref{eq8}) of} constraint (\ref{e2:c}) indicates the sleep control of SBSs, which will affect the energy consumption $\sum_{b\in \mathcal{B} } E_{b}$ in the objective. Second, the BS transmission power level $p_{b,k}$ in equation (\ref{eq7}) will affect both energy consumption and achievable rate. Finally, the achievable rate also depends on the RIS phase shift $\bm{\Theta}_{m}$, which is given by equation (\ref{eq7}). 

In summary, we aim to jointly consider the sleep control, transmission power control, and RIS phase-shift control to optimize the objective defined in equation (\ref{e2:main}).
\blue{However, note that problem (\ref{e2:main}) is highly non-convex and non-linear, involving integer control variables, fraction structure, and logarithms. 
Optimizing all variables simultaneously can be extremely difficult, especially considering the timescale differences between different control variables. 
Therefore, we decouple the joint optimization problem into two parts: a Co-HDRL algorithm is presented in the following Section \ref{s5} for sleep and power control, and then an FP-based algorithm is introduced in Section \ref{s4} for RIS phase-shift optimization.}

\section{Cooperative Hierarchical Deep Reinforcement Learning for Sleep Control and Power Control}
\label{s5}

This section introduces the proposed Co-HDRL algorithm, which applies a hierarchical architecture for decision-making, including a meta-controller for sleep control, and multiple sub-controllers for transmission power control.

\subsection{Hierarchical SMDP}
\begin{figure}[!t]
\setlength{\abovecaptionskip}{0pt} 
\centering
\subfigure[Conventional MDP architecture]{
\includegraphics[width=0.7\linewidth]{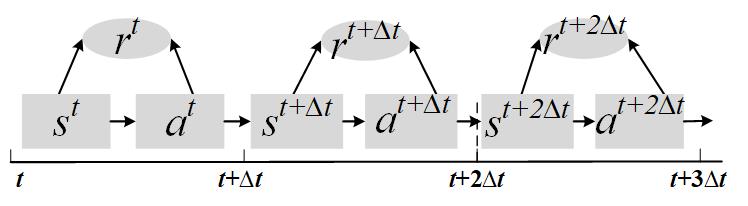} \label{fig2-a}
}
\subfigure[\textbf{\blue{SMDP architecture}}.]{
\includegraphics[width=1\linewidth]{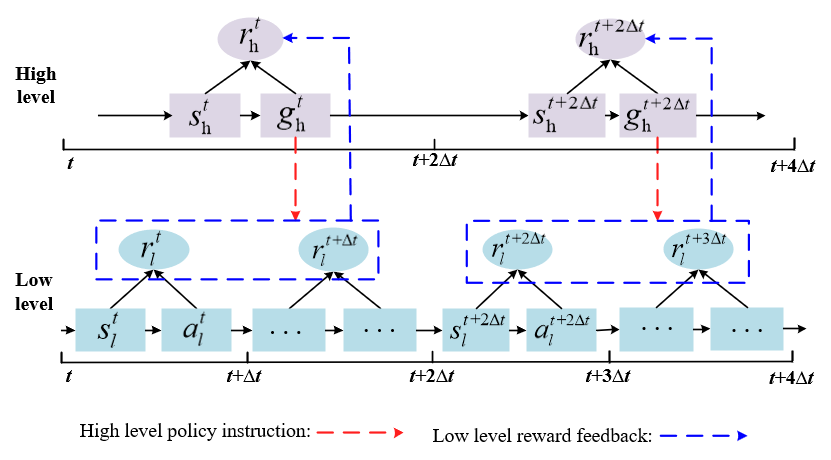} \label{fig2-b}
}
\setlength{\abovecaptionskip}{0pt} 
\caption{MDP and SMDP comparison.}
\vspace{-15pt}
\end{figure}

We first introduce a hierarchical semi Markov decision process (SMDP), which is used to transform our defined problem formulation into MDP context. 
A conventional MDP scheme can be defined by a tuple $<S,A,T,R>$, where $S$ is the state set, $A$ is the action set, $T$ is the transition probability, and $R$ is the reward function, respectively. As shown in Fig. \ref{fig2-a}, the agents select an action $a^{t}$ based on current state $s^t$, then it receives the reward $r^t$ and moves to the next state $s^{t+\Delta t}$. 
However, conventional MDP has difficulty in defining more realistic problems such as tasks with different timescale actions. For example, sleep control is a long-term decision, since turning on/off SBSs will greatly affect the network performance. By contrast, power control is a delicate and short-term decision. Therefore, we include a novel hierarchical SMDP framework to better handle the timescale difference between sleep control and power control.

Compared with conventional MDP, the main difference of SMDP lies in the hierarchical architecture. 
\blue{For example, Fig. \ref{fig2-b} includes a high-level agent and a low-level agent. 
\begin{itemize}
    \item For the high-level agent, the SMDP is defined by $<S_{H},\mathcal{G}_{H},T_{H},R_{H}>$, indicating the state, goal, transition probability, and reward of the high-level agent. The main difference compared with regular MDP is that we replace the action set $A$ by a goal set $\mathcal{G}_{H}$. Given current state $s_h^{t}$, the high-level agent will first select a goal $g_h^{t}$ at time slot $t$, and this goal serves as a high-level policy instruction for low-level agents from time slot $t$ to $t+2\Delta t$.
    \item For the low-level agent, the SMDP is defined by $<S_{L},A_{L},T_{L},R_{L},\mathcal{G}_{H}>$. It means that the low-level agents have to consider both current state $s_{l}^{t}$ and the high-level policies $g_{h}^{t}$, and then select actions $a_{l}^{t}$ accordingly. 
    Meanwhile, low-level agents can provide reward feedback to high-level agents, which can be used to evaluate high-level goals. For instance,  $r_l^{t}$ and $r_l^{t+\Delta t}$, which are low-level rewards, may affect the high-level reward $r_h^{t}$. Then the high-level agent can improve its goal selection according to the low-level reward feedback.
\end{itemize}}
\blue{These agents apply different timescales for decision making: the high-level agent selects $g_h^t$ every $2\Delta t$, and the low-level agent makes decisions every $\Delta t$.} This hierarchical timescale setting enables higher flexibility for SMDP, which makes it an ideal model to describe the joint sleep and power control problem.
Meanwhile, Fig. \ref{fig2-b} reveals that the high-level agents rely on the reward feedback of lower levels for goal evaluation. Unstable feedback from low-level agents can mislead the high-level controllers, and sub-optimal policy may be selected, which will finally degrade the overall system reward.
Note that SMDP has many variants, and here we define our SMDP to solve the hierarchical control problem \cite{b22}. 
\blue{We use two layers as an example in Fig. \ref{fig2-b}, but the proposed SMDP scheme is compatible with any number of layers.}

\begin{figure*}[!t]
\centering
\includegraphics[width=0.9\linewidth]{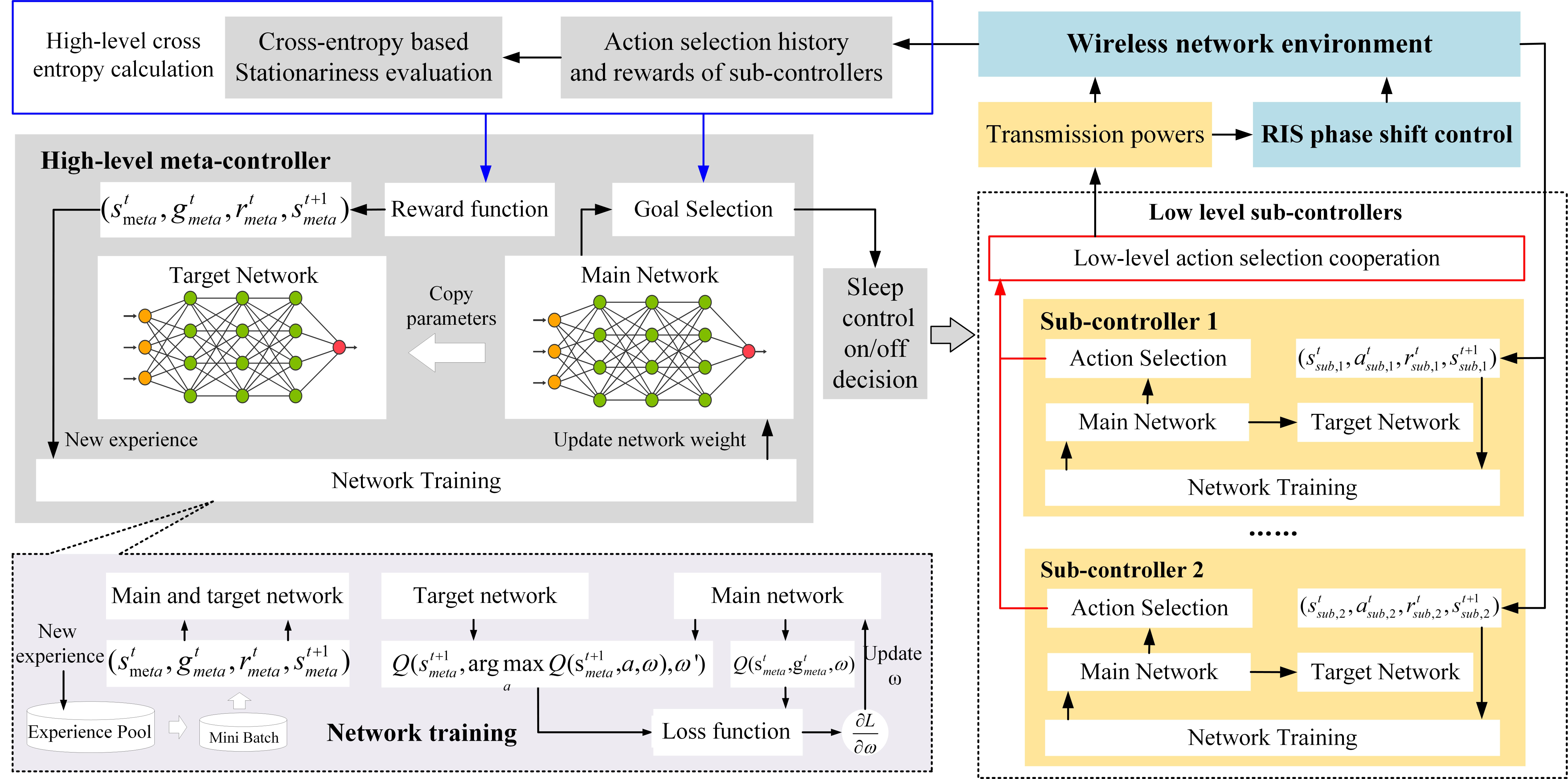}
\caption{Overall architecture of the proposed Co-HDRL.}
\label{fig3}
\vspace{-10pt}
\setlength{\abovecaptionskip}{-2pt} 
\end{figure*}

\subsection{SMDP Definition of Meta-controller and Sub-controllers}
Based on Fig. \ref{fig2-b}, we define an SMDP with two layers. The MBS is defined as a meta-controller for sleep control, which decides the long-term on/off status of attached SBSs. Then, each SBS is considered as a sub-controller to adjust its short-term transmission power level. Sleep control refers to high-level policy instructions for SBSs, and SBSs can provide reward feedback to the meta-controller to evaluate the sleep control policies.  
For the SBS sub-controller, we define the state, action, and reward by: 
\begin{itemize}
    \item \textbf{State}: For the $b^{th}$ SBS, the state $s_{sub}$ is defined by the total transmission demand level of attached UEs:  
    \begin{equation}
    s_{sub}=\frac{\sum_{k \in \mathcal{K}_{b}} W_{b,k}}{W^{max}_{b}} , \label{eq21}  
    \end{equation}
    where $W_{b,k}$ represents the transmission demand of UE $k$, $\mathcal{K}_{b}$ represents the set of UEs that are associated with $b^{th}$ SBS. $W^{max}_{b}$ is the max transmission demand of $b^{th}$ SBS, which is referred to as a constant value to normalize transmission demand in the current time slot. It is assumed the UE daily transmission demand follows specific patterns as in \cite{b23}.
    \item \textbf{Action:} The SBS sub-controller can adjust its transmission power level $p_{b,k}$ to satisfy the transmission demand in the current time slot. Therefore, the action of $b^{th}$ SBS sub-controller is $a_{sub}=\{p_{b,k}\}$ with $k \in \mathcal{K}_b $. Note that only actions that satisfy $\sum\limits_{k \in \mathcal{K}_{b}}||p_{b,k}|| \leq P_{b,max}$ will be selected, indicating the maximum transmission constraint. 
    \item \textbf{Low-level reward:} The low-level reward of sub-controller is:
    \begin{equation} \label{eq22}
    r_{sub}=\left\{ 
     \begin{array}{lcl} \frac{\sum_{k \in \mathcal{K}_{b}} C_{b,k}}{E_{b}},  &      &  \text{if\,\,} q_b=1;  \\
     \qquad 0 ,    &      &  \text{if\,\,} q_b=0;
     \end{array} \right.
     \end{equation}
    where $q_b$ has been defined in equation (\ref{eq8}) as the sleep control indicator.
    $r_{sub}$ aims to maximize the EE of $b^{th}$ SBS, and we assume the $r_{sub}$ is 0 in sleep mode. Here the $q_b$ value depends on the policy of the MBS, which means that the meta-controller can affect the reward of sub-controllers. 
\end{itemize}

The MBS is defined as a meta-controller, which will produce high-level sleep control policies for SBS sub-controllers:
\begin{itemize}
   \item \textbf{State}: The state of the MBS meta-controller includes the transmission demand level of all BSs:
    \begin{equation} \label{eq23}  
    s_{meta}=\bigg [\frac{\sum\limits_{k \in \mathcal{K}_{1}} W_{1,k}}{W^{max}_{1}}, \frac{\sum\limits_{k \in \mathcal{K}_{2}} W_{2,k}}{W^{max}_{2}},...,\frac{\sum\limits_{k \in \mathcal{K}_{|\mathcal{B}|}} W_{|\mathcal{B}|,k}}{W^{max}_{|\mathcal{B}|}}  \bigg ],
    \end{equation}
    where $\mathcal{B}$ is the set of all BSs. We consider transmission demand as states, and then the meta-controller can selectively switch off SBSs when the demand is off-peak to save overall energy consumption. 
    \item \textbf{High-level goals}: Given the transmission demand level of all BSs, the meta-controller can produce sleep control decisions for SBS sub-controllers:
    \begin{equation}
    g_{meta}=[q_{1}, q_{2},...,q_{b},...,q_{|\mathcal{B}_{SBS}|}],  \label{eq24}  
    \end{equation}
    where $q_{b}$ has been defined in equation (\ref{eq8}) as sleep mode indicator of $b^{th}$ SBS, and $\mathcal{B}_{SBS}$ is the set of SBSs.
    
   \item \textbf{High-level reward}: The meta-controller is responsible for the overall performance of all BSs. Accordingly, the high-level reward is given by the objective function we have defined in the problem formulation:
   \begin{equation}\label{eq25}
     r_{meta}=\frac{\sum_{b\in \mathcal{B} } \sum_{k \in \mathcal{K}_{b}} C_{b,k}}{\sum_{b\in \mathcal{B} } E_{b}},
   \end{equation}
      
\end{itemize}

\subsection{Co-HDRL Framework and Network Training}
The overall framework of our proposed Co-HDRL algorithm is given in Fig. 
 \ref{fig3}, which includes one meta-controller and multiple sub-controllers. The goal selected by the meta-controller will decide the on/off status of sub-controllers. 

In each controller, we deploy a double deep Q-learning (DDQN) scheme for the Q-value prediction. In conventional deep Q-learning, the loss function $Er$ of network training is:
\begin{equation}
\setlength\abovedisplayskip{5pt}
\setlength\belowdisplayskip{5pt} \label{eq26}
L(w)=Er(r^{t}+\gamma \max\limits_{a} Q(s^{t+1},a,w')-Q(s^{t},a^{t},w)),
\end{equation}
where $s^t$, $a^{t}$ and $r^t$ are the state, action and reward at time slot $t$, respectively. $w$ and $w'$ are the weight of the main and target networks. $\gamma$ is the discount factor ($0<\gamma<1$). $Q(s^{t},a^{t},w)$ is the current Q-value that is predicted by the main network, and $Q(s^{t+1},a,w')$ is the target Q-value produced by target network. 
In equation (\ref{eq26}), note that both action selection and evaluation depend on the target network, which is indicated by $\max\limits_{a} Q(s^{t+1},a,w')$. Consequently, the max operator will constantly generate overoptimistic values for the loss function, which will further affect the Q-value prediction in main network training \cite{b24}. To this end, the DDQN scheme is proposed to separate the action selection and evaluation by: 
\begin{equation}
\setlength\abovedisplayskip{3pt}
\setlength\belowdisplayskip{3pt} \label{eq27}
\begin{aligned}
L&(w)=Er(r^{t}+ \\
&\gamma Q(s^{t+1},\arg \max\limits_{a}Q(s^{t+1},a,w),w') - Q(s^{t},a^{t},w)),
\end{aligned}
\end{equation}
where $\arg \max\limits_{a}Q(s^{t+1},a,w)$ indicates the main network selects the next action, and the target network evaluates the action by $ Q(s^{t+1},\arg \max\limits_{a}Q(s^{t+1},a,w),w')$. Decoupling the action selection and evaluation can provide a more accurate target for the main network training, which will further reduce the Q-value prediction error. 

In Co-HDRL, first, the meta-controller selects a high-level goal $g_{meta}$, which indicates the sleep control decisions on sub-controllers. $g_{meta}$ is temporarily fixed in the following several time slots, and sub-controllers select actions, receive rewards, and train their networks accordingly. The transmission power of sub-controllers is sent to the RIS control block for phase shift optimization. Based on the long-term performance of sub-controllers, the meta-controller receives an average reward $r_{meta}$ from the wireless environment and moves to the next state $s_{meta}$. The new experience tuple $<s_{meta}^t,g_{meta}^t,r_{meta}^t,s_{meta}^{t+1}>$ will be sent to the experience pool (shown as the left bottom block in Fig. \ref{fig3}). Then the agent will sample a random mini batch from the experience pool, and train the main network as equation (\ref{eq27}). The target network will copy the main network weight after several training, which guarantees a stable target for the main network training.

In Fig. \ref{fig3}, it is worth noting that we define a high-level cross-entropy calculation module (shown by the blue block on the top), and a low-level action selection cooperation module (indicated by the red block on the right). In HRL, the meta-controller produces goals to instruct sub-controllers, and sub-controllers provide reward feedback to the meta-controller to evaluate the high-level goal. However, although the high-level goal $g_{meta}$ is temporarily fixed, the action of sub-controllers may be constantly changing during this period. Indeed, the sub-controllers' exploration increases the uncertainty of the reward feedback to the meta-controller. Therefore, this feedback uncertainty can mislead the goal evaluation of meta-controller\cite{b8,b16}. To this end, we proposed a cross-entropy enabled policy for high-level meta-controller, and a correlated equilibrium-based strategy for low-level sub-controllers. Following, we will introduce two techniques in detail.    

\begin{figure*}[!t]
\centering
\includegraphics[width=0.85\linewidth]{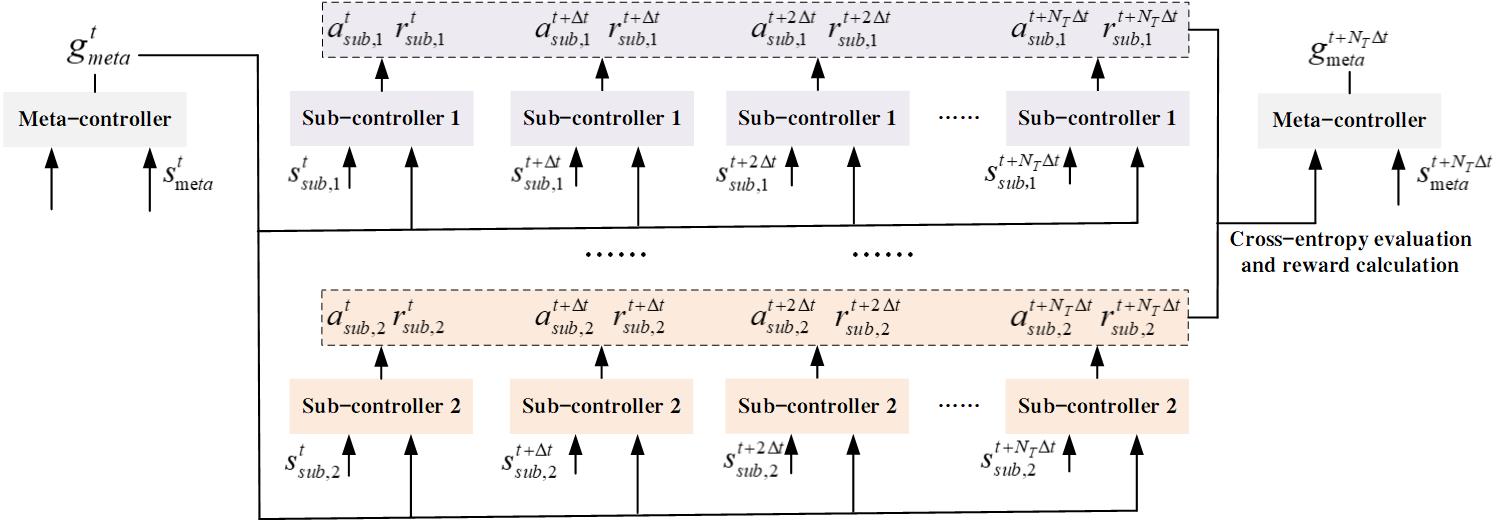}
\caption{Co-HDRL system update between meta-controller and sub-controllers.}
\label{fig4}
\setlength{\abovecaptionskip}{-2pt} 
\vspace{-10pt}
\end{figure*}

\subsection{High-level Cross-entropy enabled Meta-controller Policy}
In this subsection, we propose a cross-entropy enabled policy for meta-controller. More specifically, we use the cross-entropy as a metric to evaluate the stationarity of sub-controllers' actions, then the defined metric is used for high-level goal exploration. For example, goals with high reward feedback but low stationarity may be unreliable, since the low stationarity metric means the sub-controllers are still exploring action combinations. On the other hand, goals that bring high stationarity can be less visited in the exploration phase, because the sub-controllers have entered a stable status and require less training\cite{b25}. Following, we first present the cross-entropy metric definition, then we introduce how to use the metric for goal exploration and selection.     

Given a random variable $x$, and $N^{X}$ is the total number of possible outcomes of $x$, then the entropy of $x$ in set $X$ is defined by:
\begin{equation}
I(X)=-\sum_{i=1}^{N^{X}}pr(x_{i}) \log (x_{i}),    
\end{equation}
where $pr(x_{i})$ is the probability of $x_{i}$ in set $X$, and $\sum_{i=1}^{N^{X}}pr(x_{i})=1$. $I(X)$ indicates the average level of information and uncertainty of variable $x_{i}$ in set $X$.
Then we introduce the Kullback–Leibler divergence to define the relative entropy from one distribution $X$ to another distribution $Y$ of variable $x$ \cite{b26} :
\begin{equation} \label{eq29}
D_{KL}(X||Y)=\sum_{i=1}^{N^X}pr(x_{i})\log\frac{pr(x_{i})}{pr^{'}(x_i)}, 
\end{equation}
where $pr^{'}(x_i)$ is the probability of $x_i$ in set $Y$. $D_{KL}(X||Y)$ indicates the difference of set $X$ and $Y$ in terms of probability distribution of $x$. 
This work considers $X$ as the set of action selection history of sub-controllers, and $Y$ is the set of actions selected in the current time interval $\Delta t$. Consequently, we can apply $D_{KL}(X||Y)$ to measure the stationarity of the low-level action selection policy. The idea behind is that a stationary policy will produce similar action selections in different time intervals, which is indicated by a lower $D_{KL}(X||Y)$ value. By contrast, a higher $D_{KL}(X||Y)$ means sub-controllers require more training to stabilize the low-level policies. 

\blue{The Kullback–Leibler divergence has the minimum value when $pr(x_{i})=pr^{'}(x_{i})$, which means $D_{KL}(X||Y)=0$ if and only if $X$ and $Y$ has the same distribution $pr(x_{i})=pr^{'}(x_{i})$. A detailed proof can be found in \cite{b27} by using Gibbs' inequality.}  
Moreover, we rewrite equation (\ref{eq29}) by:
\begin{subequations}\label{e30:main}
\blue{
\begin{align}
&D_{KL}(X||Y)=\sum_{i=1}^{N^X}pr(x_{i})\log\frac{pr(x_{i})}{pr^{'}(x_i)} & \tag{\ref{e30:main}}\\
&= \sum_{i=1}^{N^X}pr(x_{i})\log(pr(x_{i}))-\sum_{i=1}^{N^X}pr(x_{i})\log(pr^{'}(x_{i})) & \nonumber\\
&= -I(X)-\sum_{i=1}^{N^X}pr(x_{i})\log(pr^{'}(x_{i})), & \nonumber
\end{align}}
\end{subequations}
where $I(X)$ is the entropy of the action selection distribution history, and $Y$ is the action selection distribution in the current time interval. \blue{The history distribution $X$ indicates the action selection probability of all previous $t$ time slots, while $Y$ represents the probability of current time slot $\Delta t$. Therefore, the history distribution $X$ usually changes slowly as a long-term metric, and $Y$ may change quickly since the action selection of the current time interval may change frequently due to agent exploration. Therefore, it means the history distribution $X$ is generally more stable compared with the current distribution $Y$. In equation (\ref{e30:main}), $\sum_{i=1}^{N^X}pr(x_{i})\log(pr^{'}(x_{i}))$ item contributes more to the uncertainty than $I(x)$, which is known as the cross-entropy\cite{b28}.} In this work, we use cross entropy to represent the stationarity of the sub-controllers. 

This work includes multiple sub-controllers, and the cross-entropy of $b^{th}$ sub-controller under high-level goal $g_{meta}$ is:
\begin{equation}\label{eq31}
\blue{I(X_{b,t-1},Y_{b,t}|g_{meta})=-\sum_{i=1}^{A_{g_{meta}}}pr(a_{i})\log(pr^{'}(a_{i}))},    
\end{equation}
where $X_{b,t-1}$ is the accumulated action selection distribution of sub-controllers in former $t-1$ time slots under high-level goal $g_{meta}$, $Y_{b,t-1}$ is the action selection distribution under $g_{meta}$ in current time slot, and $A_{g_{meta}}$ is the action set of sub-controllers under $g_{meta}$. The cross-entropy metric \blue{$I(X_{b,t-1},Y_{b,t}|g_{meta})$} defines the stationary of sub-controller action selections under $g_{meta}$.  
Finally, in the exploration phase, the meta-controller selects high-level goals by: 
\begin{equation}\label{eq32}
\blue{\begin{aligned}
pr(g_{meta}|s_{meta})= \qquad \qquad \qquad \qquad  \qquad   \qquad \qquad\\ 
\frac{\tanh(\sum_{b\in \mathcal{B}_{-M}}I(X_{b,t-1},Y_{b,t}|g_{meta}))}{\sum\limits_{g_{meta}\in \mathcal{G}}\tanh(\sum_{b\in \mathcal{B}_{-M}}I(X_{b,t-1},Y_{b,t}|g_{meta}))},  
\end{aligned}}
\end{equation}
where $\mathcal{B}_{-M}$ indicates the set of controllers except the MBS meta-controller. The $\tanh$ indicates the Tanh function, which is applied to normalize all the cross-entropy values.

\begin{algorithm}[!t] 
	\caption{Co-HDRL based joint sleep and power control}
	\begin{algorithmic}[1] \label{a2}
		\STATE \blue{\textbf{Input:} Wireless network parameters: BS, user, and RIS locations, RIS element numbers, BS power consumption level, Rician channel settings, traffic models; Co-HDRL algorithm settings: experience pool size, training frequency, minibatch size, neural network architecture and learning rate.}
		 \FOR{every $N_T\Delta t$, the MBS}
		 \STATE Updating the cross-entropy metric under $g_{meta}$ based on the action selection history of last $N_T\Delta t$.  
         \STATE Calculating the average EE metric of former $N_T\Delta t$ as high-level reward $r_{meta}$ by equation (\ref{e2:main}). 
		 \STATE Updating new state $s_{meta}$, and saving $(s_{meta},g_{meta},$ $r_{meta},s_{meta}')$ to the experience pool. 
		 \STATE \blue{Sampling a mini-batch from the experience pool randomly. Then generate the target Q-value $Q^{Tar}(s_{meta},g_{meta})=r_{meta}$ if done; otherwise $Q^{Tar}(s_{meta},g_{meta})=r_{meta}+ \gamma Q(s_{meta}',\arg \max\limits_{g}$ $Q(s_{meta}',g,w),w').$}
        \STATE Updating $w$ using gradient descent by minimizing the loss $L(w)=Er(Q^{Tar}(s_{meta},g_{meta})-Q(s_{meta},g_{meta},w))$.
		\STATE Copying $w$ to $w'$ after several training.
		\STATE Selecting a high-level goal $g_{meta}$ of the next $N_T\Delta t$ by following method: 
		 \STATE with probability $\epsilon$, selecting high-level goal $g_{meta}$ by equation (\ref{eq32}); otherwise, choosing goal $g_{meta}$ by greedy policy $\arg \max\limits_{a}Q(s_{meta},g_{meta},w)$.
		\ENDFOR
		\FOR{every $\Delta t$, active SBSs}
		 \STATE Selecting the transmission power level $a_{sub}$ by: 
		 \STATE with probability $\epsilon$, sub-controllers select $a_{sub}$ randomly; otherwise, choosing joint action $\vec a_{sub}$ by correlated equilibrium as equation (\ref{eq33:main}).
		 \STATE Sending the transmission power to RIS control algorithm, then receives the data transmission rate and calculates the low-level reward.
		 \STATE Updating new state $s_{sub}$, and saving $(s_{sub},a_{sub},$ $r_{sub},s_{sub}')$ to the experience pool. Each sub-controller samples a mini-batch from its experience pool, and trains the network similar to the meta-controller. 
		\ENDFOR
	\STATE \blue{\textbf{Output:} Network performance metrics: Power consumption, data rate, and EE of each BS; Algorithm learning performance: rewards of meta-controller and sub-controllers, and system cross-entropy values defined by equation (\ref{eq31})}.	
	\end{algorithmic}
\end{algorithm}

\begin{algorithm}[!t] 
	\caption{HDRL-based joint sleep and power control}
	\begin{algorithmic}[1] \label{a4}
	  \STATE \blue{\textbf{Input:} Wireless network parameters: BS, user, and RIS locations, RIS element numbers, BS power consumption level, Rician channel settings, traffic models; Co-HDRL algorithm settings: experience pool size, training frequency, minibatch size, neural network architecture and learning rate.}
		 \FOR{every $N_T\Delta t$, the MBS}
         \STATE Calculating the average EE metric of former $N_T\Delta t$ as high-level reward $r_{meta}$ by equation (\ref{e2:main}). 
		 \STATE Updating new state $s_{meta}$, and saving $(s_{meta},g_{meta},$ $r_{meta},s_{meta}')$ to the experience pool. 
		 \STATE Sampling a mini-batch from experience pool randomly. Generating target Q-values \begin{small}$Q^{Tar}(s_{meta},g_{meta})$= 
             $\left\{
             \begin{array}{ccl}
            r_{meta} &   if\;done\\
              r_{meta}+ \max \gamma Q(s_{meta}',g_{meta},w') &  else
            \end{array} \right.$\end{small}
        \STATE Updating $w$ using gradient descent by minimizing the loss $L(w)=Er(Q^{Tar}(s_{meta},g_{meta})-Q(s_{meta},g_{meta},w))$.
		\STATE Copying $w$ to $w'$ after several training.
		\STATE With probability $\epsilon$, selecting a high-level goal randomly for the next $N_T\Delta t$; otherwise, choosing goal $g_{meta}$ by greedy policy $\arg \max\limits_{a}Q(s_{meta},g_{meta},w)$.
		\ENDFOR
		\FOR{every $\Delta t$, active SBSs}
		 \STATE Selecting the transmission power level $a_{sub}$ by: 
		 \STATE with probability $\epsilon$, sub-controllers select $a_{sub}$ randomly; otherwise, choosing action $a_{sub}$ by greedy policy.
		 \STATE Sending the transmission power to RIS control algorithm, then receives the data transmission rate and calculates the low-level reward. 
		 \STATE Updating new state $s_{sub}$, and saving $(s_{sub},a_{sub},$ $r_{sub},s_{sub}')$ to the experience pool. Each sub-controller samples a mini-batch from its experience pool, and trains the network similar to the meta-controller. 
		\ENDFOR
	\STATE \blue{\textbf{Output:} Network performance metrics: Power consumption, data rate, and EE of each BS; Algorithm learning performance: rewards of meta-controller and sub-controllers.}.
	\end{algorithmic}
\end{algorithm}

In summary, equation (\ref{eq32}) shows that goals that lead to lower stability will be more frequently selected in the exploration phase, which is indicated by a high cross-entropy value. It guarantees that sub-controllers can provide reliable and stable feedback to the meta-controller for goal evaluation. \blue{The meta-controllers and sub-controllers are updated as Fig. \ref{fig4}. In particular, the parameters associated with each sub-controller are updated in a distributed manner, which has been presented in previous Fig. \ref{fig3}. For instance, each sub-controller will receive the high-level goal $g_{meta}$ and environment state $s_{sub}$ and then select action $a_{sub}$ autonomously. Such a distributed optimization approach will reduce the communication overhead between sub-controllers, and it can also overcome the computational burden caused by a single central-controller approach\cite{xu2023algorithm}.}
The meta-controller will first select a goal $g^{t}_{meta}$ at time $t$, which is temporarily fixed in the next $N_T\Delta t$ time slot. Then, sub-controllers select the actions in each $\Delta t$ under current state $s_{sub}$ and high-level goal $g^{t}_{meta}$. Finally, the actions and rewards of sub-controllers from $t$ to $t+N_T\Delta t$ are collected by meta-controllers for cross-entropy calculation and goal selection as shown by equation (\ref{eq32}).

\subsection{Low-level Sub-controllers Cooperation} 
We propose a correlated equilibrium-based cooperation strategy for low-level sub-controllers. Correlated equilibrium is proposed as a multi-agent collaboration strategy\cite{b12}. Compared with existing methods such as Nash equilibrium, the existence and convergence of correlated equilibrium can be better guaranteed by linear programming solutions. Indeed, cooperative action selections can bring higher stability than independent action selections. With correlated equilibrium, the SBS sub-controllers choose actions by:   
\begin{subequations}
\label{eq33:main}
\begin{align}
\max_{a_{sub}\in A_{sub}} \sum_{\vec a_{sub} \in A_{sub}} &pr(\vec a_{sub}|s_{sub})  Q(s_{sub},\vec a_{sub},w) \tag{\ref{eq33:main}}   \\
\blue{s.t.} & \quad 0 \leq pr(\vec a_{sub}|s_{sub})\leq 1, \label{eq33:a}\\
&\sum_{\vec a \in A}pr(\vec a_{sub}|s_{sub})=1, \label{eq33:b}\\
\sum_{a_{-b}\in A_{-b} }&pr(\vec a_{sub}|s_{sub})(Q(s_{sub},\vec a_{sub},w)& \nonumber\\
& -Q(s_{sub},\vec a_{-b}, a_{b},w))\geq0, \label{eq33:c}
\end{align}
\end{subequations}
where $pr(\vec a_{sub}|s_{sub})$ is the probability of selecting action combination $\vec a_{sub}=(a_{1},a_{2},...,a_{|\mathcal{B}_{-M}|})$ under state $s_{sub}$, $\vec a_{-b}$ indicates the joint action of all sub-controllers except $b^{th}$ sub-controller, and $w$ is the main network weight. $\mathcal{B}_{-M}$ has been defined in equation (\ref{eq32}) as the set of controllers except the MBS meta-controller, $A_{-b}$ is the set of $\vec a_{-b}$. 

Equation (\ref{eq33:main}) uses Q-values predicted by the main network to represent the expected reward. It aims to maximize the total potential reward of all sub-controllers by finding an optimal action selection probability distribution $pr(\vec a_{-b}|s_{sub})$.
Constraints (\ref{eq33:a}) and (\ref{eq33:b}) are probability distribution constraints.
The constraint (\ref{eq33:c}) indicates that $\vec a_{-b}$ produces higher expected reward with probability $pr(\vec a_{-b}|s_{sub})$, which is indicated by $pr(\vec a_{sub}|s_{sub})(Q(s_{sub},\vec a_{sub},w) -Q(s_{sub},\vec a_{-b}, a_{b},w))\geq0$.

Finally, the proposed Co-HDRL algorithm is summarized in Algorithm \ref{a2}. \blue{The algorithm inputs include wireless network parameters such as BS, user, and RIS locations, number of RIS elements, BS power consumption levels, etc, and the Co-HDRL algorithm settings involve the neural network learning rate and training frequency, experience pool size, and so on. The algorithm outputs consist of network performance metrics such as power consumption, data rate, and EE of each BS, and meanwhile the learning performance metrics such as rewards and cross-entropy values.}

\subsection{Computational Complexity Analyses} 
This section analyzes the computational complexity of the proposed algorithms. Neural network training contributes to the major runtime complexity in Co-HDRL. In this work, we deploy Long short-term memory (LSTM) networks for the Q-values prediction, which is a special recurrent neural network that can better capture the long-term dependency of input data. The computational complexity of updating LSTM networks is $O(\alpha^2 \chi^2)$, where $\alpha$ indicates the memory blocks number, and $\chi$ represents the memory cell number of one block \cite{b24}.

\subsection{Baseline: Hierarchical Deep Reinforcement Learning}
To better demonstrate the capability of our proposed Co-HDRL algorithm, we propose a conventional hierarchical deep reinforcement learning (HDRL) based method as a benchmark.  \blue{The HDRL is given in Algorithm \ref{a4}, which has similar inputs as Co-HDRL in Algorithm \ref{a2}.} In HDRL, both meta-controller and sub-controllers apply $\epsilon$-greedy policy for goal and action selections, respectively, and there is no high-level cross-entropy and low-level cooperation between agents. 
\blue{Therefore, compared with Algorithm \ref{a2}, the outputs of Algorithm \ref{a4} do not include cross entropy metrics defined by equation (\ref{eq31}).}

\section{RIS Phase-shift Optimization} \label{s4}
\blue{This work decouples the EE maximization problem into sleep control, power control, and RIS phase-shift optimization problems. While Co-HDRL has been presented in the above Section \ref{s5} for sleep and power control, this section will solve RIS phase-shift optimization problems. In particular, an FP-based approach is applied to optimize RIS phase shifts in Section \ref{sec-fp}. Meanwhile, considering the high design complexity of FP approaches, we also proposed a low-complexity baseline algorithm in Section \ref{surrogate}, namely the surrogate optimization-based method.}

\subsection{Fractional Programming based method}
\label{sec-fp}
There have been many advanced algorithms for RIS phase-shift design, and here we select the FP method because: 1) FP is a straightforward method that is widely used in many existing studies; and 2) this work focuses on hierarchical learning for decision-making of different timescales, and it is reasonable to apply a well-known technique for RIS phase-shift optimization\cite{zhou2023survey}.
The objective of RIS phase-shift control is to maximize the total data rate for all UEs:
\begin{equation}\label{e9:main}
\resizebox{0.67\hsize}{!}{$\begin{aligned}
\max\limits_{\bm{\Theta}_{m}}  & f_{1}(\bm{\Theta}_{m})=\sum_{k\in \mathcal{K}}b_{k} \log(1+\psi_{k})  \\
 \text{s.t.} \quad &|\theta_{m,n}|^2=1, m\in\mathcal{M}, n\in\mathcal{N}_m,
\end{aligned}$}
\end{equation}
where \blue{$\psi_{k}$ is the SINR that can be extracted from equation (\ref{eq7}) and $\psi_{k}=\frac{|\sum\limits_{m\in \mathcal{M}}\bm{H}_{b,m}\bm{\Theta}_{m}\bm{G}_{m,k}^{\dag}p_{b,k}|^2}{\sum\limits_{b'\in \mathcal{B}} \sum\limits_{k'\in \mathcal{K}_{b'},k'\neq k}|\sum\limits_{m'\in \mathcal{M}}\bm{H}_{b',m'}\bm{\Theta}_{m'}\bm{G}_{m',k}^{\dag}p_{b',k'}|^2+N_{0}^2}$}. 
According to \cite{b9}, problem (\ref{e9:main}) is equivalent to:

\begin{equation}\label{e10:main}
\blue{
\begin{aligned}
\max\limits_{\bm{\Theta}_{m},\bm{\beta}} f_{2}(\bm{\Theta}_{m},\bm{\beta}) = 
\sum_{k\in \mathcal{K}} \bigg (b_{k} \log(1+\beta_{k}) - \qquad \qquad  \\ 
 b_{k}\beta_{k}+\frac{b_{k}(1+\beta_{k})\psi_{k}}{1+\psi_{k}} \bigg )\\  
\text{s.t.}  \quad |\theta_{m,n}|^2=1, m\in\mathcal{M}, n\in\mathcal{N}_m, 
\end{aligned}}
\end{equation}
where $\bm{\beta}=[\beta_1,\beta_2,...,\beta_{|\mathcal{K}|}]$ is the auxiliary variables given by Lagrangian dual transform. 
To solve problem (\ref{e10:main}), we apply an iterative method to update the $\bm{\Theta}_{m}$ and $\bm{\beta}$ alternatively. First, given $\bm{\Theta}_{m}$, setting $\frac{\partial f_{2}}{\partial \beta_{k}}=0$ claims $\beta^*_{k}=\psi_{k}$. Then, given $\beta^*_{k}$, optimizing $\bm{\Theta}_{m}$ means:
\blue{
\begin{equation}\label{e11:main}
\begin{aligned}
\max\limits_{\bm{\Theta}_{m}} f_{3}(\bm{\Theta}_{m})= \qquad \qquad \qquad \qquad \qquad \qquad \qquad\\ 
\sum_{k\in \mathcal{K}} \frac{b_{k}(1+\beta_{k})|\sum\limits_{m\in \mathcal{M}}\bm{H}_{b,m}\bm{\Theta}_{m}\bm{G}_{m,k}^{\dag}p_{b,k}|^2}{\sum\limits_{k'\in \mathcal{K}_b}|\sum\limits_{m'\in \mathcal{M}}\bm{H}_{b',m'}\bm{\Theta}_{m'}\bm{G}_{m',k}^{\dag}p_{b',k'}|^2+N_{0}^2}\\
 \text{s.t.}  \quad |\theta_{m,n}|^2=1, m\in\mathcal{M}, n\in\mathcal{N}_m, \qquad \qquad \qquad 
\end{aligned}
\end{equation}}

For ease of notation, we define $\bm{\hat{\theta}}_{m}=[\theta_{m,1},\theta_{m,2}, ...,$ $\theta_{m,\mathcal{N}_m}]$,
and $\sqrt{P_{b}}\bm{H}_{b,m}\bm{\Theta}_{m}\bm{G}_{m,k}^{\dag}$ can be easily transformed to $\omega_{m}\bm{\hat{\theta}}_{m} $\text{diag}$(\bm{H}_{b,m})\bm{G}_{m,k}^{\dag}\sqrt{P_{b}}$. For notation brevity, we further define $\bm{v}_{b,m,k}=\omega_{m}\text{diag}(\bm{H}_{b,m})\bm{G}_{m,k}^{\dag}\sqrt{P_{b}}$. Based on quadratic transformation\cite{b9}, equation (\ref{e11:main}) is equivalent to (the proof is given in the appendix):
\begin{equation}\label{e13:main}
\resizebox{0.8\hsize}{!}{$\begin{aligned}
\max\limits_{\bm{\hat{\Theta}},\eta_{k}} \ & f_{4}(\bm{\hat{\Theta}},\eta_{k})= \sum_{k\in \mathcal{K}}( 2\sqrt{b_{k}(1+\beta_{k})}\\ 
& \qquad  \text{Re}\{\eta_{k}^{\dag}(\bm{\hat{\Theta}} \bm{V}_{b,k})\}-\eta_{k}^{\dag}(\sum\limits_{b'\in \mathcal{B}} |\bm{\hat{\Theta}} \bm{V}_{b',k}|^2+N_{0}^2)\eta_{k}) \\
  \text{s.t.}  \quad &|\theta_{m,n}|^2=1, m\in\mathcal{M}, n\in\mathcal{N}_m, 
\end{aligned}$}
\end{equation}
where $\bm{\hat{\Theta}}=[\bm{\hat{\theta}}^{\dag}_{1},\bm{\hat{\theta}}^{\dag}_{2},...,\bm{\hat{\theta}}^{\dag}_{|\mathcal{M}|}]$ represents all the RIS phase control variables, $\bm{V}_{b,k}=[\bm{v}_{b,1,k},\bm{v}_{b,2,k},...,\bm{v}_{b,|\mathcal{M}|,k}]$, $\bm{\eta}=[\eta_{1},\eta_{2},...\eta_{|\mathcal{K}|}]$ a collection of auxiliary variables, and $\text{Re}\{\}$ refers to the real number part of a complex number.    

Then, we rewrite equation (\ref{e13:main}) to optimize $\bm{\hat{\Theta}}$: 
\begin{equation}\label{e15:main}
\resizebox{0.8\hsize}{!}{$\begin{aligned}
\max\limits_{\bm{\hat{\Theta}}} \ & f_{5}(\bm{\hat{\Theta}})= -\bm{\hat{\Theta}^{\dag}} \Lambda \bm{\hat{\Theta}}+2 \text{Re}\{\bm{\hat{\Theta}^{\dag}}\Psi \}+\sum_{k\in \mathcal{K}}|\eta_{k}|^2 N_{0}^2 \\
  \text{s.t.}  \quad &|\theta_{m,n}|^2=1, m\in\mathcal{M}, n\in\mathcal{N}_m, 
\end{aligned}$}
\end{equation}
where $\Lambda=\sum_{k\in \mathcal{K}} |\eta_{k}|^2 \sum\limits_{b'\in \mathcal{B}}|\bm{V}_{b',k}|^2$ and $\Psi=\sum_{k\in \mathcal{K}} 2$ $\sqrt{b_{k}(1+\beta_{k})}\eta_{k}^{\dag} \bm{V}_{b,k}$ for ease of notations. Equation (\ref{e15:main}) is a quadratically constrained quadratic programming problem, but the phase shift constraint $|\theta_{m,n}|^2=1$ is non-convex. Therefore, we relax this constraint by allowing $|\theta_{m,n}|^2\leq1$ for convexity, which means:
\begin{equation}\label{e16:main}
\begin{aligned}
\max\limits_{\bm{\hat{\Theta}}} \ & f_{6}(\bm{\hat{\Theta}})= -\bm{\hat{\Theta}^{\dag}} \Lambda \bm{\hat{\Theta}}+2 \text{Re}\{\bm{\hat{\Theta}^{\dag}}\Psi \} \\
 \text{s.t.}  \ & |\theta_{m,n}|^2\leq1, m\in\mathcal{M}, n\in\mathcal{N}_m,
\end{aligned}
\end{equation}
which is an optimization problem with strong convexity. The Lagrange dual of this problem can be written as an unconstrained problem:
\begin{equation}\label{e17:main}
\resizebox{0.88\hsize}{!}{$\begin{aligned}
\max\limits_{\bm{\hat{\Theta}},\bm{\hat{\sigma}}} \quad & f_{7}(\bm{\hat{\Theta}},\bm{\hat{\sigma}})= f_{6}(\bm{\hat{\Theta}}) -\sum\limits_{m\in\mathcal{M}}\sum\limits_{n\in\mathcal{N}_m} \sigma_{m,n} (|\theta_{m,n}|^2-1),
\end{aligned}$}
\end{equation}
where $\sigma_{m,n}$ is the Lagrange dual variable for each constraint with $\bm{\hat{\sigma}}=[\bm{\sigma}_{1},...,\bm{\sigma}_{m},...,\bm{\sigma}_{\mathcal{M}}]$, and each 
$\bm{\sigma}_{m}=[\sigma_{m,1},...,$ $\sigma_{m,n},...,\sigma_{m,|\mathcal{N}_m|}]$. 
Setting $\frac{\partial f_{7}}{\partial \bm{\hat{\Theta}}}=0$, then we have the optimal  $\bm{\hat{\Theta}^*}=\frac{\Psi}{\Lambda+\text{diag}(\bm{\hat{\sigma}})}$.
Given $\bm{\hat{\Theta}^*}$, setting $\frac{\partial f_{7}}{\partial \sigma_{m,n}}=0$ will bring $|\theta_{m,n}|^2=1$, which means each element in $\bm{\hat{\Theta}^*}$ will satisfy the constraints.  
Therefore, the optimal value of equation (\ref{e16:main}) can be achieved by updating optimal $\bm{\hat{\Theta}}$ and $\bm{\hat{\sigma}}$ iteratively. 
Finally, we optimize three control variables, $\beta$, $\eta$, and $\bm{\hat{\sigma}}$ alternatively, which is known as the alternating optimization approach. This means optimizing one variable at each time and holding other variables fixed, then it moves to the next variable until the solutions converge\cite{zhou2023survey}. 

\subsection{\blue{Baseline: Surrogate Optimization based RIS phase control}}
\label{surrogate}
This subsection introduces the proposed surrogate optimization-based RIS phase-shift control.
In practice, the phase shift can only be selected from a set of fixed values, which mainly depends on the predefined resolution settings of phase shifters. 
Different than the FP-based method, the surrogate optimization treats the optimization as a black-box problem\cite{b31}. It yields a reasonable approximation for the original optimization problem by a surrogate model, which avoids the complexity of designing a dedicated optimization solution.
Therefore, we formalize the RIS control problem to maximize the total transmission rate: 
\blue{
\begin{equation}\label{eq29-s}
\begin{aligned}
\max\limits_{\mathcal{Z}} &f_{8}(\mathcal{Z})=\sum_{k\in \mathcal{K}}b_{k} \log \bigg(1+ \psi_{k} \bigg)  \\
 \text{s.t.} \ & \theta^{'}_{m,n} =z_{m,n}\frac{2\pi}{2^\mu},  \\
 & z_{m,n} \in \left\{0,1,\cdots, (2^\mu -1) \right\},
\end{aligned}
\end{equation}}
where $\theta^{'}_{m,n}$ is the RIS phase-shift angle. $z_{m,n}$ belongs to the matrix $\mathcal{Z}$, which will decide the phase shift angle $\theta^{'}_{m,n}$, \blue{and $\psi_{k}$ is the SINR defined in equation (\ref{e9:main}).} Note that the decision variable $z_{m,n}$ can only be selected from fixed integer values, which means equation (\ref{eq29-s}) is a Mixed Integer Nonlinear Programming (MINLP) problem, then a surrogate optimization method is applied to \blue{optimize} this problem. \blue{However, despite the low design complexity, there are no guarantees of the algorithm performance since MINLP problems are NP-hard.} 
\blue{There are various approaches to design surrogate optimization algorithms, and here we consider a serial algorithm.
In particular, it includes the surrogate construction phase and minimum searching phase, which relies on random samples to construct and update a surrogate of unknown objective functions \cite{MATLAB}. 
}


\blue{In summary, we propose a Co-HDRL algorithm in the previous Section \ref{s5} for joint sleep and power control, and then an FP-based method in Section \ref{s4} for RIS phase-shift optimization. 
Co-HDRL applies a reinforcement learning-enabled approach to improve the decision quality by iteratively interacting with the environment, while the convex optimization-based FP algorithm focuses on problem transformation to achieve convexity. 
Despite the differences, note that the meta-controller in Co-HDRL will first make sleep control decisions as high-level policies, and then sub-controllers can change the transmission power.
Given the sleep and power control decisions, the FP algorithm is implemented for RIS phase-shift optimization, calculating the channel capacity. 
Therefore, the RIS optimization module can be considered as part of the environment reward function, generating the reward based on action input. It indicates that the hierarchical reinforcement learning framework dominates the whole system, and the optimality aligns with the reinforcement learning scheme.  
}

\begin{table}[!tb]
\caption{Parameters settings}
\centering
\renewcommand\arraystretch{1.4}
\begin{tabular}{|p{3.7cm}<{\centering}||p{4.1cm}<{\centering}|}
\hline
 \textbf{Wireless network settings} & \textbf{Traffic model}\\
\hline
3GPP urban macro network & 20\% Poisson distribution \\ 
Bandwidth: 20MHz & 80\% constant bit rate \\ 
\cline{2-2}
Number of RBs: 100 & \textbf{Learning and algorithm settings} \\
\cline{2-2}
  Subcarriers in each RB: 12& Network layers: 4 \\
  Subcarrier bandwidth: 15kHz & 2 LSTM network as hidden layers  \\
  \cline{1-1}
\textbf{UE and BSs settings} & Hidden layer has 10 nodes \\
\cline{1-1}
1 MBS, 3 SBSs  &  Experience pool size: 600  \\
MBS cell radius: 400m  & Initial network learning rate: 0.005 \\
SBS cell radius: 100m  &  Minibatch size: 180 \\
MBS max Tx power: 20 W  & Discount factor: 0.3 \\
SBS max Tx power: 6.3 W  & Epsilon value: 0.05 \\
20 UE with random distribution & Training frequency: 240 iterations \\
 \cline{1-1}
 \textbf{RIS settings} & Network copy: every 600 iterations \\
 \cline{1-1}
 6 RIS deployed in the cell & Time resolution difference $N_{T}$:10 \\
 Each RIS includes 10 elements & Overloading penalty factor: 0.2\\
\hline
\end{tabular}
\label{tab2}
\end{table} 

\section{Performance Evaluation} \label{s7}

\subsection{Simulation Settings}

We consider a network that includes 1 MBS and 3 SBSs. The radius of the MBS and SBS are 400 m and 100 m, respectively. The carrier frequency for the MBS and SBS are both 4 GHz\cite{elsayed2019reinforcement}. The load-dependent power consumption factor is 4.7 and 2.6 for MBS and SBS, respectively\cite{b23}. The whole cell includes 20 UEs that are randomly distributed. 6 RISs are uniformly distributed, and each contains 10 elements with 2 bits resolution. 
For learning settings, we apply 2 LSTM networks as hidden layers for each controller. The hyperparameters such as neural network learning rate and training frequency are selected by the grid search method, which means we try different parameter combinations and find the best results accordingly. 
The simulations are repeated 10 runs on MATLAB with 95\% confidence interval, and other settings are summarized in Table \ref{tab2}.

\subsection{Performance Comparison under Sleep Control and RIS}
\begin{figure}[!t]
\centering 
\subfigure[Power consumption of RIS and no-RIS cases]{ \includegraphics[width=0.75\linewidth]{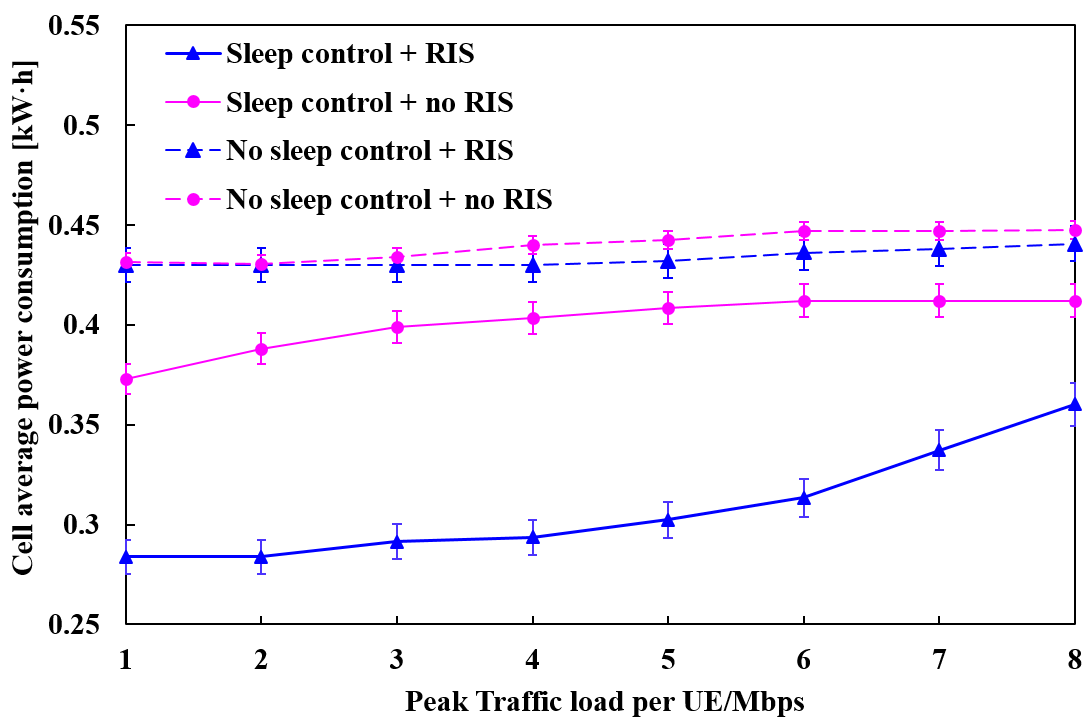} \label{f-r1-1}
}
\subfigure[Energy efficiency comparison of RIS and no-RIS cases]{
\includegraphics[width=0.75\linewidth]{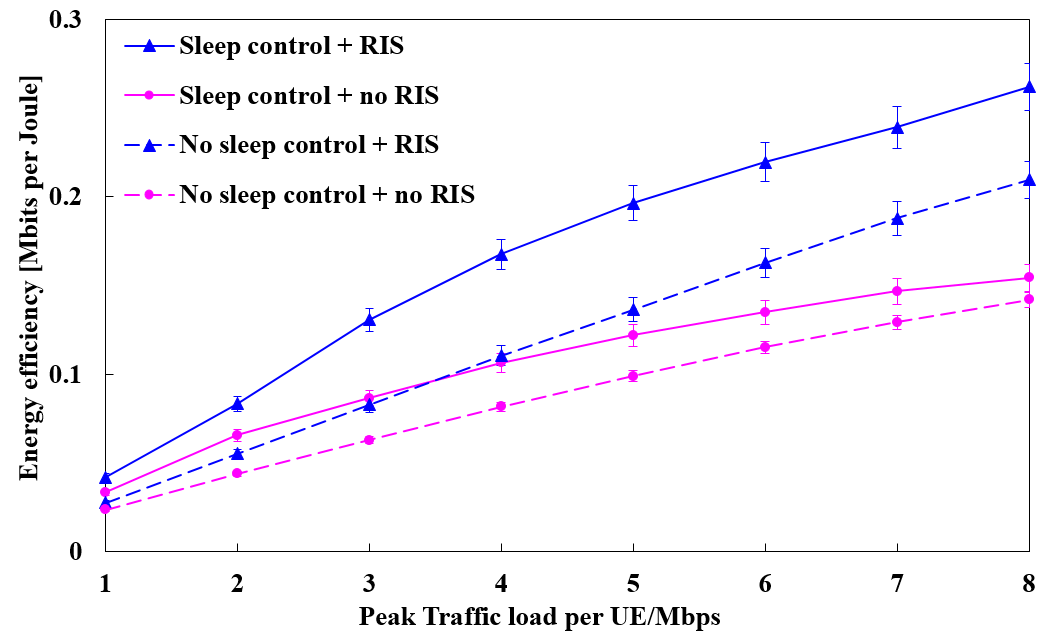}\label{f-r1-3}
}
\subfigure[Daily SBS on/off status analyses]{
\includegraphics[width=0.75\linewidth]{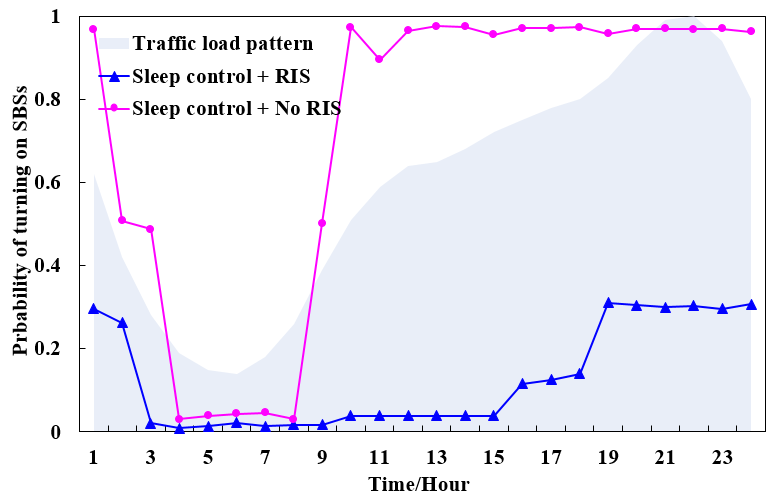} \label{f-r1-4}
}
\caption{\textbf{\blue{Performance comparison of sleep control and RIS.}}}
\label{f-r1}
\vspace{-20pt}
\end{figure}

This section investigates the performance under sleep control and RIS techniques. 
The proposed FP and Co-HDRL methods are deployed for RIS control and sleep control, respectively. As shown in Fig. \ref{f-r1}, four cases are considered by combining sleep control with RIS control. 
\blue{Here we focus on the relationship between sleep control and RISs because combining transmission power and RISs to improve EE has been widely studied in existing works \cite{zhou2023survey}, while the relationship between RISs and sleep control has not been well investigated in previous studies \cite{wu2015energy}.}

Fig. \ref{f-r1-1} shows that sleep control can significantly save power consumption by switching off SBSs intelligently. Combining sleep control with RIS can achieve the lowest energy consumption. 
\blue{Fig. \ref{f-r1-3} demonstrates that combining sleep control with RISs can achieve higher EE than other combinations. 
This is because RISs can increase the channel capacity of the signal transmission, and it means that more SBSs can enter sleep mode to save energy consumption, which is also demonstrated in Fig. \ref{f-r1-4}. 
Meanwhile, deploying RISs or implementing sleep control independently will produce comparable EE performance. In particular, using RISs can improve the channel capacity for higher throughput, while sleep control can greatly save power consumption in off-peak periods. Then these two scenarios present comparable EE results. 
By contrast, the no sleep control + no RIS case shows the lowest EE as a regular baseline approach. 
}
We further compare the daily SBS on/off status of RIS and no-RIS condition in Fig. \ref{f-r1-4}. The vertical axis is the probability of turning on SBSs, which indicates the power consumption level during specific time slots. The blue shade represents the preset traffic load pattern. 
\blue{Fig. \ref{f-r1-4}} shows that most SBSs are switched off during the low-peak period (from 3:00 to 7:00) in both RIS and no-RIS cases. However, when the traffic load increases to mid-peak and on-peak periods, the no-RIS case has to turn on most SBSs to satisfy the increasing traffic demand; otherwise, the overloading may lead to a high penalty. By contrast, using RIS can greatly increase the channel capacity, which means the MBS can still switch off most SBSs to save energy without causing overloading.    
\blue{To summarize, Fig. \ref{f-r1} demonstrates that combining sleep control and RISs can achieve higher EE than using two techniques separately.}

\begin{figure*}[!t]
\centering
\subfigure[Convergence performance of surrogate optimization based RIS control ]{
\includegraphics[width=5.7cm,height=4.1cm]{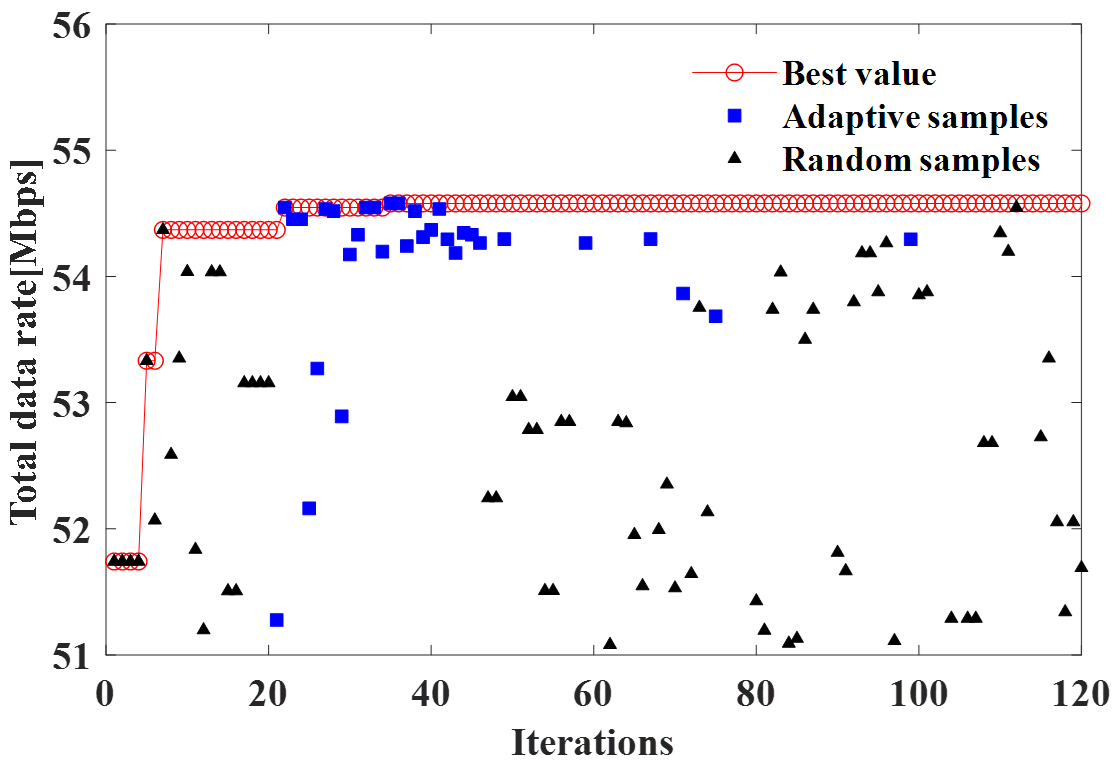} \label{f-r2-1}
}
\subfigure[Convergence performance of fractional programming (FP) based RIS control]{
\includegraphics[width=5.7cm,height=4.1cm]{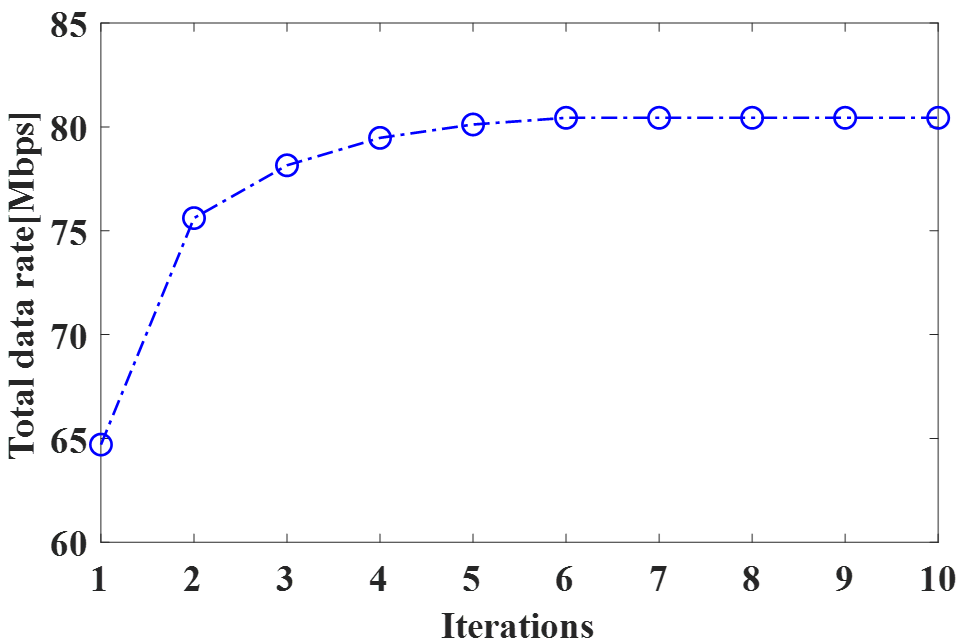} \label{f-r2-2}
}
\subfigure[Power consumption comparison of different RIS control methods ]{
\includegraphics[width=5.7cm,height=4.1cm]{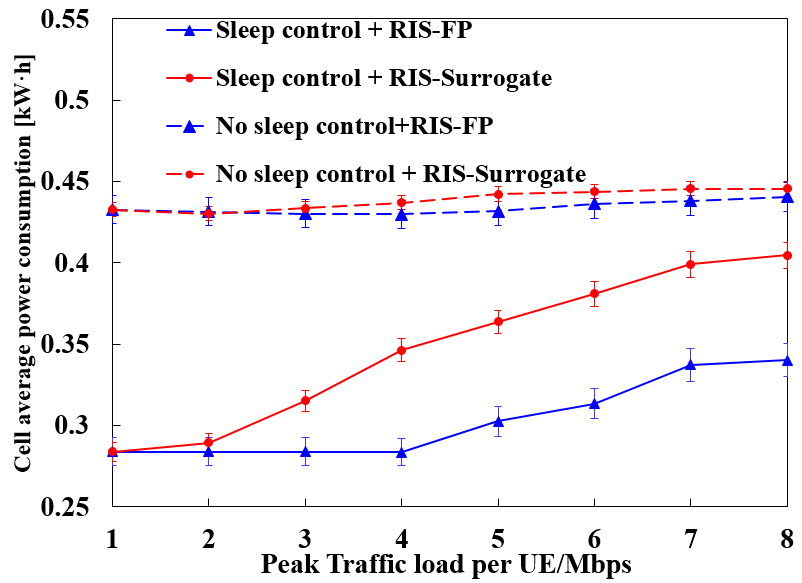} \label{f-r2-3}
}
\quad
\subfigure[Throughput comparison of different RIS control methods]{
\includegraphics[width=5.7cm,height=4.1cm]{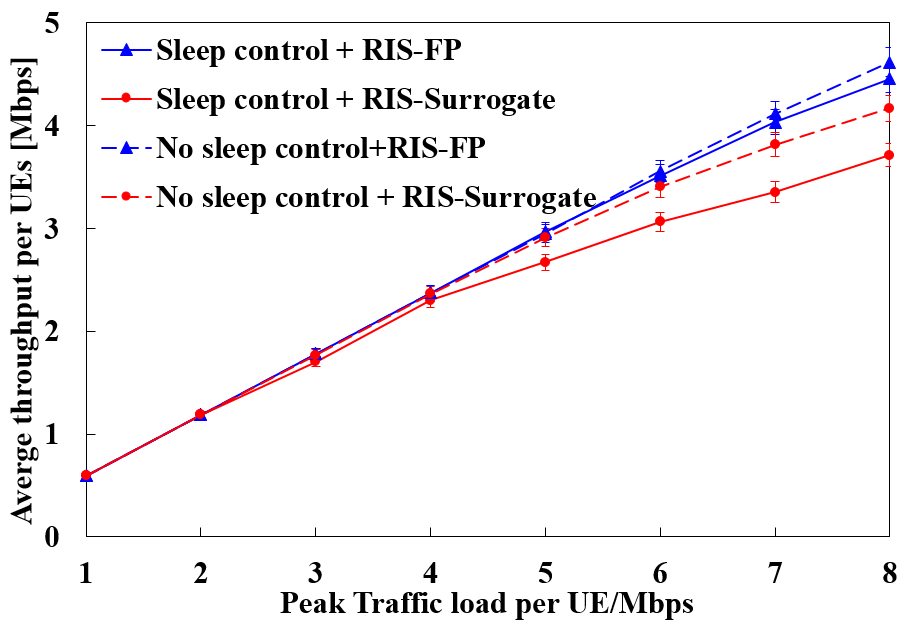} \label{f-r2-4}
}
\subfigure[ Energy efficiency comparison of different RIS control methods]{
\includegraphics[width=5.7cm,height=4.1cm]{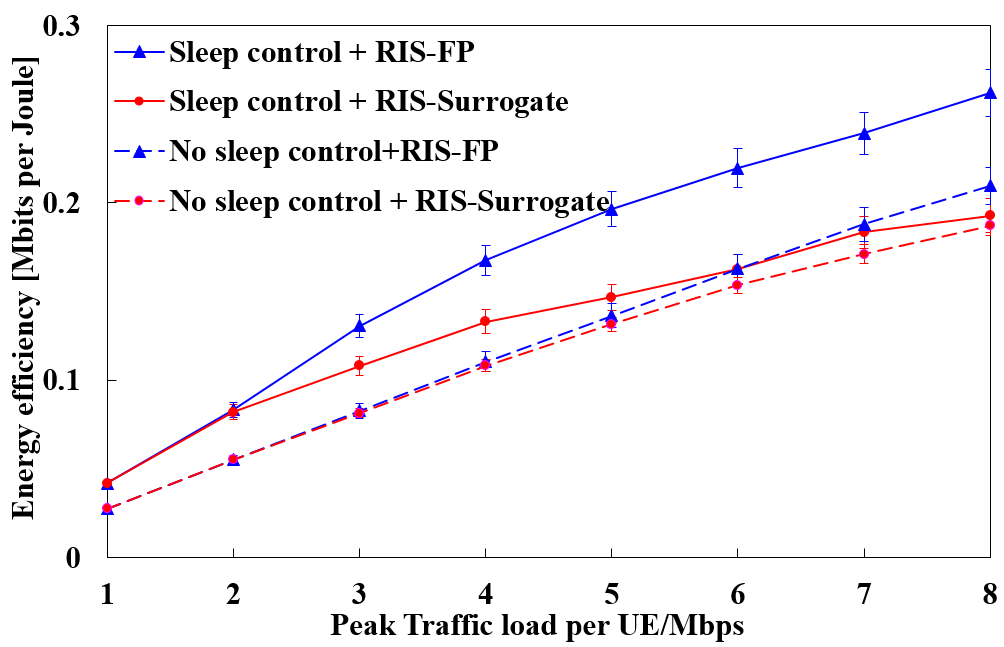} \label{f-r2-5}
}
\subfigure[Daily SBS on/off status analyses of different RIS control methods]{
\includegraphics[width=5.7cm,height=4.1cm]{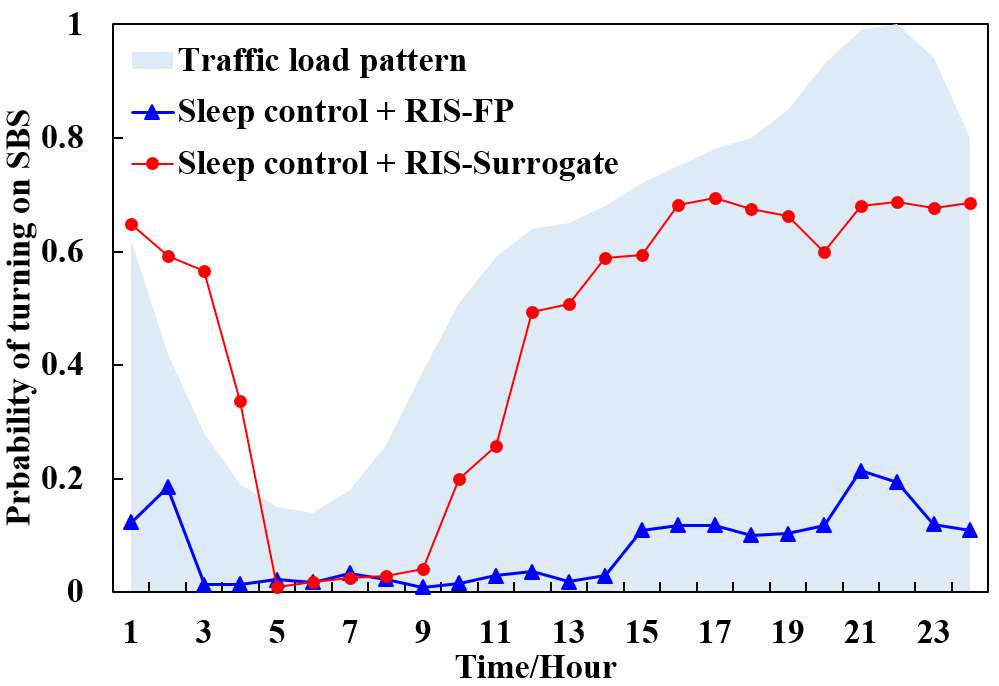} \label{f-r2-6}
}
\caption{Performance comparison of difference RIS control strategies.}
\label{f-r2}
\vspace{-13pt}
\end{figure*}

\subsection{RIS Control Strategy Analyses}
This subsection compares the proposed FP-based RIS control method with the surrogate optimization strategy.  
Firstly, Fig. \ref{f-r2-1} shows the convergence performance of surrogate optimization in one episode. 
\blue{This work applies a serial algorithm to construct the surrogate functions, including surrogate construction and minimum searching phases  \cite{MATLAB}.  
One can observe that surrogate optimization relies on random and adaptive samples to search for better solutions, which are shown by black and blue dots in Fig. \ref{f-r2-1}. Then the surrogate model is constantly updated in each iteration to provide a more accurate estimation of the given problem.} 
By contrast, in Fig. \ref{f-r2-2}, our proposed FP method reaches a much faster convergence. This faster convergence can be explained by the dedicated designed FP optimization model for RIS control. In the proposed FP, the alternative optimization method guarantees that the next iteration can always perform better than the former iterations.

Similarly, we consider the combination of sleep control with different RIS control methods, and the power consumption, throughput and EE results are given in Fig. \ref{f-r2-3}, (d) and (e), respectively. Combining FP-based RIS control with sleep control can bring lower power consumption, higher throughput and EE than other cases. The proposed FP method can achieve 35\% lower power consumption and 16\% higher EE when the peak traffic load is 8 Mbps. 
%
Moreover, the SBS on/off status is shown in Fig. \ref{f-r2-6}, in which FP-based RIS control can maintain low power consumption even in peak-load periods (from 17:00 to 23:00). On the contrary, surrogate-based RIS control has to turn on SBSs to adapt to the increasing transmission demand to prevent overloading penalty. 
\blue{In summary, Fig. \ref{f-r2} shows the importance of designing appropriate RIS phase-shift optimization techniques to improve the system EE performance. Compared with surrogate optimization, the FP-based method can better control the phase shifts of RIS elements, allowing more SBSs to enter sleep mode to save energy consumption and therefore obtain higher system EE. }

\begin{figure}[!t]
\centering
\subfigure[Energy efficiency comparison of Co-HDRL and HDRL]{
\includegraphics[width=0.75\linewidth]{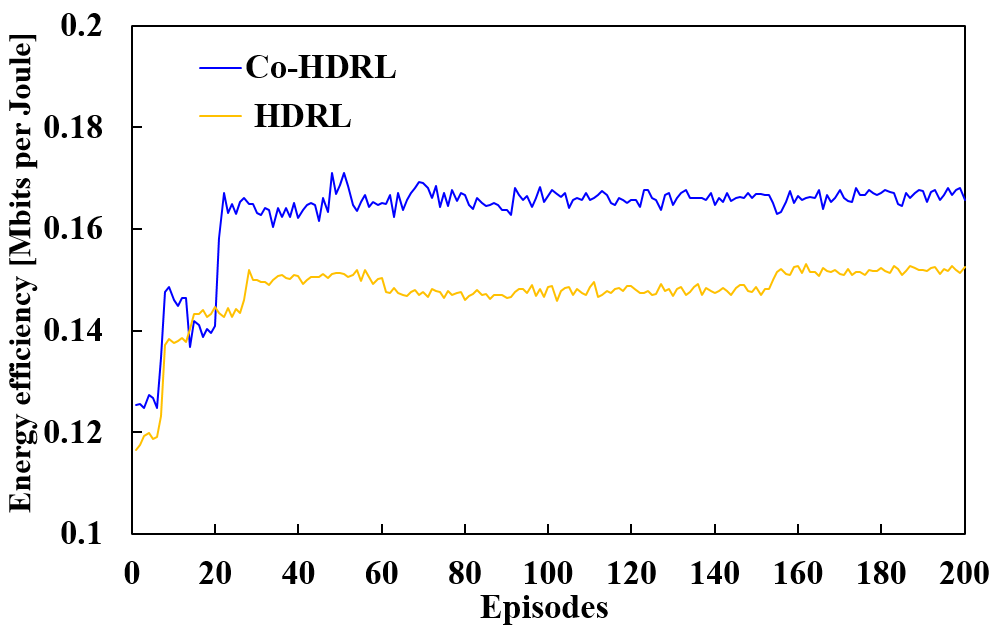} \label{f-r3-1}
}
\subfigure[Stationarity comparison of different action selection strategies]{
\includegraphics[width=0.75\linewidth]{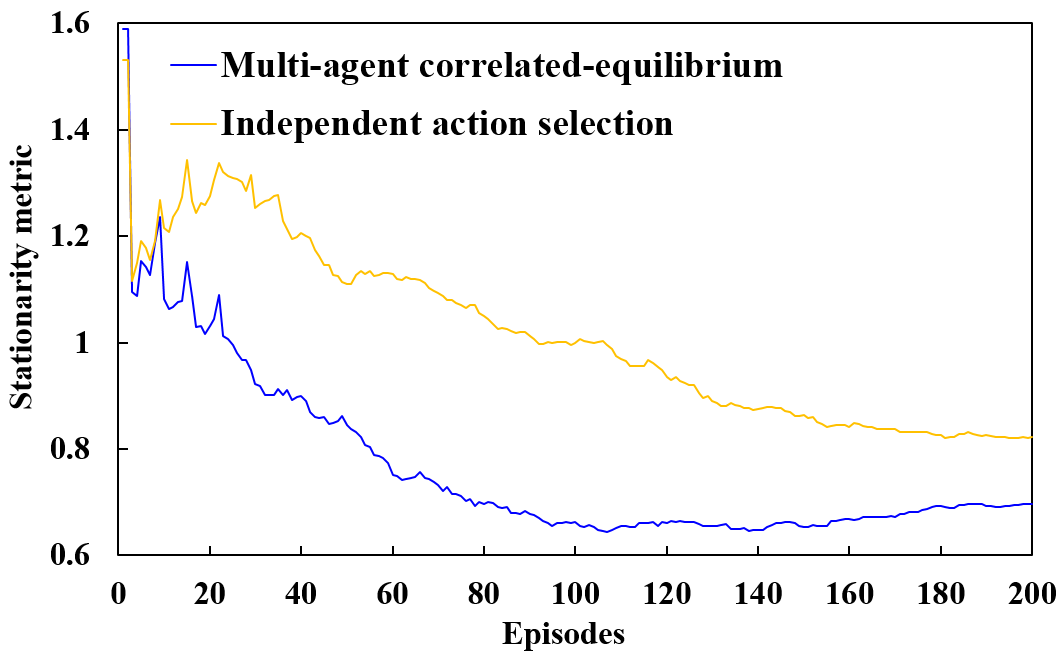} \label{f-r3-3}
}
\quad
\subfigure[Daily SBS on/off status of Co-HDRL and HDRL]{
\includegraphics[width=0.75\linewidth]{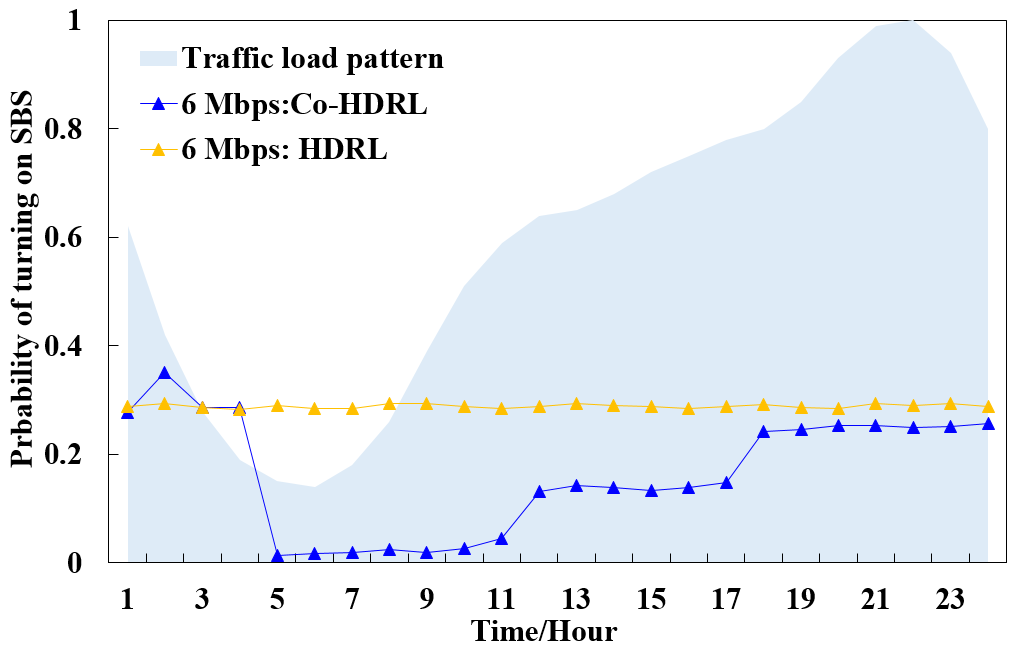} \label{f-r3-4}
}
\subfigure[ Daily SBS on/off status of Co-HDRL under various traffic load]{
\includegraphics[width=0.75\linewidth]{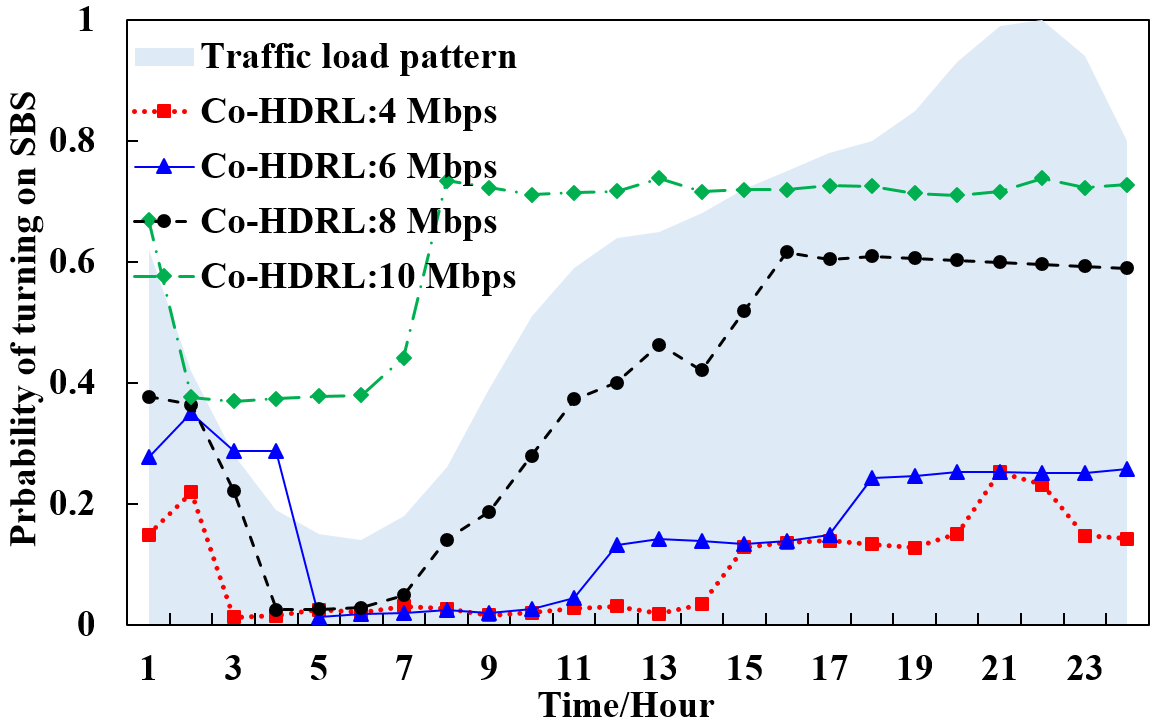} \label{f-r3-5}
}
\caption{\textbf{\blue{Performance analyses of Co-HDRL and HDRL.}}}
\label{f-r3}
\vspace{-13pt}
\end{figure}

\subsection{Machine Learning Algorithm Comparison}

In this section, we compare the proposed Co-HDRL algorithm with conventional HDRL, which aims to prove that the cooperation strategy can bring better performance. \blue{The EE performance of 200 episodes is given in Fig. \ref{f-r3-1}, in which the proposed Co-HDRL algorithm achieves a higher EE and reward than the HDRL baseline.}
\blue{Specifically, the proposed Co-HDRL algorithm applies the cross-entropy-based policy for the meta-controller to monitor the stability of sub-controllers, and the correlated equilibrium is deployed for joint action selection of sub-controllers. These cooperation methods enable higher stability for sub-controller action selection and training, which brings higher EE than conventional HDRL.} 
In Fig. \ref{f-r3-3}, we further present the stationarity of various low-level action selection strategies. The proposed multi-agent correlated equilibrium enables a more stable action selection than independent action selection that is applied in HDRL, which is indicated by a lower stationarity metric. Meanwhile, Fig. \ref{f-r3-3} demonstrates that correlated equilibrium can reach a stable state much faster than the independent action selection method.   
Meanwhile, the satisfying performance of Co-HDRL can also be observed in Fig. \ref{f-r3-4}, in which the Co-HDRL switches on/off SBSs dynamically during on-peak and off-peak periods. By contrast, HDRL produces the same sleep control decision regardless of different traffic load conditions ( traffic patterns shown by the light blue shade). Moreover, the Co-HDRL performance from 4 to 10 Mbps is presented in Fig. \ref{f-r3-5}, in which the proposed Co-HDRL make decisions intelligently under various traffic load level. 

\blue{Previous simulations and results have demonstrated the benefits of applying Co-HDRL to RIS-aided energy-efficient networks. In the future, we plan to investigate the application of hierarchical learning schemes in more advanced metasurface techniques, e.g., stacked intelligent metasurfaces \cite{an2023stacked, an2024stacked}, which can take full advantage of the proposed Co-HDRL algorithms.}

\section{Conclusion} \label{s8}
Machine learning is a promising technique for network management in 5G and future 6G networks. In this work, we propose a cooperative hierarchical deep reinforcement learning method for RIS-assisted energy efficient RAN, including a cross-entropy enabled meta-controller policy and cooperative action selections for sub-controllers. Besides, a fractional programming-based RIS control strategy is introduced for phase shift control. The proposed architecture enables joint sleep, power, and RIS control with different timescales, increasing the flexibility for RAN management. This is particularly important for 6G and O-RAN where intelligence will be more abundant and control will span several time scales.
The proposed method is compared with conventional hierarchical deep reinforcement learning and surrogate optimization-based RIS control, and the simulations show that the proposed algorithms achieve better energy efficiency. In the future, we will focus more on the RIS control methods, and investigate how the RIS location affects the network performance.

\appendices
 \section{Proof}
\label{appendix-B}
From equation (\ref{e11:main}), we have 
\begin{equation}\label{e12:main} \nonumber
\resizebox{0.8\hsize}{!}{$\begin{aligned}
\max\limits_{\bm{\hat{\Theta}}} \ & f_{3}(\bm{\hat{\Theta}}) =\sum_{k\in \mathcal{K}} \frac{b_{k}(1+\beta_{k})|\sum\limits_{m\in \mathcal{M}} \bm{\hat{\theta}}^{\dag}_{m} \bm{v}_{b,m,k}|^2 }{\sum\limits_{b'\in \mathcal{B}}|\sum\limits_{m'\in \mathcal{M}}\bm{\hat{\theta}^{\dag}}_{m'} \bm{v}_{b',m',k}|^2+N_{0}^2} \\
& \qquad = \sum_{k\in \mathcal{K}} \frac{b_{k}(1+\beta_{k})|\bm{\hat{\Theta}} \bm{V}_{b,k}|^2 }{\sum\limits_{b'\in \mathcal{B}}|\bm{\hat{\Theta}} \bm{V}_{b',k}|^2+N_{0}^2}  \\
\text{s.t.} \quad &|\theta_{m,n}|^2=1, m\in\mathcal{M}, n\in\mathcal{N}_m,
\end{aligned}$}
\end{equation}
where $\bm{\hat{\Theta}}=[\bm{\hat{\theta}}^{\dag}_{1},\bm{\hat{\theta}}^{\dag}_{2},...,\bm{\hat{\theta}}^{\dag}_{|\mathcal{M}|}]$ represents all the RIS phase control variables, and $\bm{V}_{b,k}=[\bm{v}_{b,1,k},\bm{v}_{b,2,k},...,\bm{v}_{b,|\mathcal{M}|,k}]$. 
For ease of notations, we use $A_{k}$ to represent $ \sqrt{b_{k}(1+\beta_{k})} (\bm{\hat{\Theta}} \bm{V}_{b,k})$, and $B_{k}$ to represent $\sum\limits_{b'\in \mathcal{B}} |\bm{\hat{\Theta}} \bm{V}_{b',k}|^2+N_{0}^2$. 
Then, we can rewrite equation (\ref{e13:main}) as $2\text{Re}\{ \eta_{k}^{\dag} A_{k}\}-\eta_{k}^{\dag}B_{k}\eta_{k}$, which can be reformulated by $ \eta_{k}^{\dag} A_{k}+A_{k}^{\dag}\eta_{k}-\eta_{k}^{\dag}B_{k}\eta_{k}$. Then we rewrite it as $A_{k}^{\dag}B^{-1}_{k}A_{k}-(\eta_{k}^{\dag}-B^{-1}_{k}A_{k}^{\dag})B_{k}(\eta_{k}-B^{-1}_{k}A_{k})$. It is obvious that the optimal value is $\eta^*_{k}=B^{-1}_{k}A_{k}$, which means 
\begin{equation}\nonumber
  \eta^*_{k}= \frac{\sqrt{b_{k}(1+\beta_{k})}\bm{\hat{\Theta}} \bm{V}_{b,k}}{\sum\limits_{b'\in \mathcal{B}} |\bm{\hat{\Theta}} \bm{V}_{b',k}|^2+N_{0}^2}.  
\end{equation}
Then substituting $\eta^*_{k}$ back to equation (\ref{e13:main}) will get the optimal value as $\sum_{k\in \mathcal{K}} \frac{b_{k}(1+\beta_{k})|\bm{\hat{\Theta}} \bm{V}_{b,k}|^2 }{\sum\limits_{b'\in \mathcal{B}}|\bm{\hat{\Theta}} \bm{V}_{b',k}|^2+N_{0}^2}$, which is exactly the objective of the former problem. As such, the equivalence is demonstrated.

\bibliographystyle{IEEEtran}
\bibliography{Globecom2022}

\end{document}